\documentclass[twocolumn,twocolappendix]{aastex631}
\usepackage{newtxtext,newtxmath}

\begin{document}

\title{Coherent Inverse Compton Scattering in Fast Radio Bursts Revisited}

\author[0000-0003-4721-4869]{Yuanhong Qu}\thanks{E-mail: yuanhong.qu@unlv.edu}
\affiliation{Nevada Center for Astrophysics, University of Nevada, Las Vegas, NV 89154}
\affiliation{Department of Physics and Astronomy, University of Nevada Las Vegas, Las Vegas, NV 89154, USA}

\author[0000-0002-9725-2524]{Bing Zhang}\thanks{E-mail: bing.zhang@unlv.edu}
\affiliation{Nevada Center for Astrophysics, University of Nevada, Las Vegas, NV 89154}
\affiliation{Department of Physics and Astronomy, University of Nevada Las Vegas, Las Vegas, NV 89154, USA}

\begin{abstract}
Growing observations of temporal, spectral, and polarization properties of fast radio bursts (FRBs) indicate that the radio emission of the majority of bursts is likely produced inside the magnetosphere of its central engine, likely a magnetar. We revisit the idea that FRBs are generated via coherent inverse Compton scattering (ICS) off low-frequency X-mode electromagnetic waves (fast magnetosonic waves) by bunches at a distance of a few hundred times of the magnetar radius. Following findings are revealed: 1. Crustal oscillations during a flaring event would excite kHz Alfv\'en waves. Fast magnetosonic waves with essentially the same frequency can be generated directly or be converted from Alfv\'en waves at a large radius, with an amplitude large enough to power FRBs via the ICS process.  2. The cross section increases rapidly with radius and significant ICS can occur at $r \gtrsim 100 R_\star$ with emission power much greater than the curvature radiation power but still in the linear scattering regime. 3. The low-frequency fast magnetosonic waves naturally redistribute a fluctuating relativistic plasma in the charge-depleted region to form bunches with the right size to power FRBs. 4. The required bunch net charge density can be sub-Goldreich-Julian, which allows a strong parallel electric field to accelerate the charges, maintain the bunches, and continuously power FRB emission. 
5. This model can account for a wide range of observed properties of repeating FRB bursts, including high degrees of linear and circular polarization and narrow spectra as observed in many bursts from repeating FRB sources. 
\end{abstract}


\section{Introduction}
Fast radio bursts (FRBs) are bright radio pulses detected in the frequencies from $\sim 100$ MHz to several GHz with extremely high brightness temperatures ($\sim10^{35}$ K), which demand an extremely high degree of coherence of the radiation. Large excessive dispersion measures with respect to the Milky Way values suggest a cosmological origin \citep{Lorimer2007,Thornton2013,Spitler2014,petroff2019}, and some FRB sources have been well localized in galaxies at cosmological distances \citep{Chatterjee2017,Tendulkar2017,Bannister2019,Prochaska2019,Ravi2019,Macquart2020,Marcote2020,Niu2022,Xu2021}. The detection of the FRB-like event, FRB 200428 \citep{Bochenek2020,CHIME/FRB2020} in association with an X-ray burst \citep{CKLi21,Mereghetti20} from the Galactic magnetar SGR 1935+2154 suggested that at least some FRBs can be produced from magnetars. Even though the sources of cosmological FRBs are not identified, it is widely believed that magnetars have the required physical properties to account for most of them. 
The energy source to power FRBs may come from various forms, e.g. magnetic, strain, rotational or gravitational \citep[e.g.][]{Wang2024}. 
However, the radiation site (inside or outside the magnetosphere) and the radiation mechanism are largely unknown and still subject to debate.

\begin{figure*}
\includegraphics[width=18.5 cm,height=10 cm]{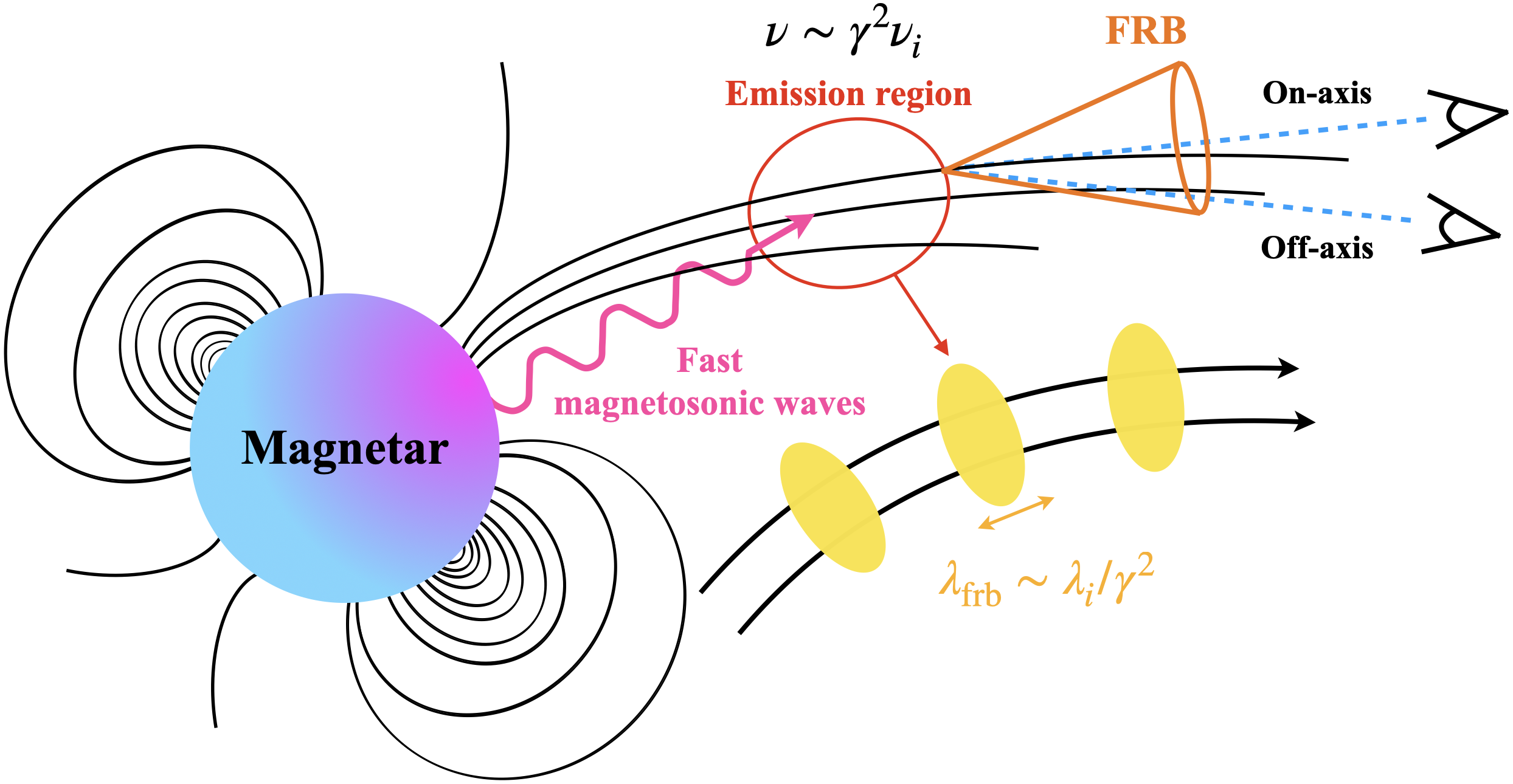}
    \caption{The geometric sketch of the magnetospheric coherent ICS model. 
    The low-frequency fast magnetosonic waves (purple wiggler) can be generated near the neutron star surface (or converted from Alfv\'en waves) with an effective near-surface electric field amplitude of $E_{\rm fmw,0}\sim10^8-10^{11}$ esu. Relativistic bunched particles (zoomed in and marked in yellow) in the emission region upscatter the low-frequency waves to power $\sim$ GHz FRB emission. 100\% linearly polarized waves and high circularly polarized waves can be observed along the on-axis and off-axis directions, respectively. The longitudinal size of the charged bunch is $\sim\lambda_{\rm frb}\sim\lambda_i/\gamma^2$ and such fluctuations of plasma density with respect to the background are formed due to modulation by the incident fast magnetosonic waves. The separations between bunches are uniform so that emission from multiple bunches is also superposed coherently to reach a narrow emission spectrum.}
    \label{fig:cartoon}
\end{figure*}

Within the magnetar framework, one can generally classify radiation models to two classes based on the location where coherent radiation is emitted \citep{Zhang2020,Lyubarsky2021,ZhangRMP}: pulsar-like models (close-in models) that invoke emission processes inside or slightly outside the magnetospheres \citep[e.g.][]{Kumar2017,Yang&zhang2018,Wang2019,Yang&zhang2021,Wadiasingh19,kumar&Bosnjak2020,Lu20,Lyutikov2021,Zhang22,QKZ,Liu22} 
and GRB-like models (far-away models) that invoke emission processes in relativistic shocks far from the magnetospheres 
\citep[e.g.][]{Lyubarsky2014,Beloborodov2017,Beloborodov2020,Plotnikov&Sironi2019,Metzger2019,Margalit2020}. 
Both types of models can account for some FRB properties but encounter difficulties or criticisms either from theoretical or observational perspectives. In general, growing observational evidence supports the magnetospheric origin of the majority of bursts \citep[e.g.][]{Zhang2023Universe}, which we summarize below: 
\begin{itemize} 

\item Polarization angle (PA) swings have been clearly observed at least in some FRBs \citep{Luo2020nature}, with at least one case consistent with the ``S"-shaped as predicted by the pulsar rotation vector model \citep{Mckinven2024}. All these are consistent with a magnetospheric origin with line of sight sweeping across different magnetic field lines. There was an attempt to accommodate PA swing in the shock model by invoking O mode generation \citep{Iwamoto2023}, with the expense of losing linear polarization degree and radiation efficiency, inconsistent with the observations.

\item Significant circular polarization has been detected in some FRBs, both
repeating \citep{CHIME2021,Xu2021,ZhouDJ2022,ZhangYK2023,Jiang22,NiuJR2022,Feng2022b} and non-repeating  \citep{masui2015,petroff2015,Cho2020,Day2020,Sherman2023}.
In principle, both intrinsic radiation mechanism and propagation effects can induce high circular polarization \citep{Qu&Zhang2023}. Some sources (e.g. FRB 20220912A) reside in a clean environment with rotation measure close to zero but still carry strong circular polarization, which favors an intrinsic magnetospheric radiation mechanism \citep{ZhangYK2023,Feng2022}.

\item It has been known that repeating FRBs have narrower spectra than non-repeating ones \citep{Pleunis2021}. Very narrow spectra with a relative width $\Delta\nu/\nu_0\sim0.2-0.3$ (the bandwidth $\Delta\nu$ is defined at the full width at the half-maximum (FWHM) and $\nu_0$ is the central frequency) have been observed in many bursts from some repeating sources  \citep{ZhouDJ2022,ZhangYK2023,Sheikh2024}. 
Theoretically, a generic constraint $\Delta\nu/\nu_0\gtrsim0.58$ ({for FWHM}) due to the high-latitude effect can be posed on any jet model that invokes a wide conical jet geometry \citep{KQZ2024}. This is particular challenging for the shock models, but is reconcilable for magnetospheric models that invoke narrow emission beams defined by magnetic field lines. 
{The synchrotron maser model might be responsible for some non-repeating FRBs with broad frequency bandwidth.}

\item Very rapid variability with a tens-of-nanosecond variability timescale has been detected in some bursts \citep{Nimmo2022,Snelders2023}, which posed challenges to the shock models but is consistent with the magnetospheric origin of FRBs \citep{Beniamini&Kumar2020}.
\end{itemize}

These key observational results call for the development of a magnetospheric emission model that can satisfy all the observational constraints\footnote{The theoretical arguments used to disfavor FRB emission from magnetospheres \citep{Beloborodov2021,Beloborodov2022} do not apply to realistic magnetospheric FRB emission models that invoke open field lines and relativistic particles. Under these realistic situations, there is no problem for FRBs being generated and propagating in magnetospheres \citep{QKZ,Lyutikov2024}.}. The most widely discussed model invokes coherent curvature radiation (CR) by bunches \citep{Katz2018,Kumar2017,Yang&zhang2018,Lu2020,Cooper2021,Wang2022,Wang2022b,Yang&Zhang2023}. This mechanism, inherited from the radio pulsar models \citep{Ruderman1975}, is user friendly and has well-defined features. It, however, suffers a list of criticisms (some inherited from the pulsar field): 1. The most severe theoretical difficulty is the excitement of maintenance of the bunches with the right size \citep{Melrose1978}. Even though two-stream instabilities have been invoked to make bunches, it is contrived to make bunches with the size comparable to the wavelength of FRB emission. 2. Within the context of FRBs, in order to sustain a high power of relativistic particles to make FRB emission, a strong parallel electric field $E_\parallel$ is needed in the emission region, which requires significant charge depletion \citep{Kumar2017}. On the other hand, a relatively large net charge factor with respect to the Goldreich Julian \citep{Goldreich&Julian1969} density is required to reach the desired coherence level. This requires a high-density plasma, which tends to screen $E_\parallel$ in the emission region (\citealt{Lyubarsky2021}, but see \citealt{QZK}). 3. Recent observations show that the emission spectra of repeating FRBs are extremely narrow \citep{ZhouDJ2022,ZhangYK2023}. CR has an intrinsically wide spectrum. Even though charge separation \citep{Yang2020} or spatially properly distributed bunches \citep{Yang2023,Wang2023} may make spectra narrower, but it is not straightforward how the required spatial distribution can be realized.

\cite{Zhang22} proposed an alternative coherent mechanism by bunches by invoking the inverse Compton scattering (ICS) process to interpret FRBs. The ingredients of the model include a type of low-frequency X-mode electromagnetic waves (likely fast magnetosonic waves) induced by neutron star crust oscillations during the FRB triggering events and relativistic particles in a charge depleted region where $E_\parallel$ continues to pump energy to the relativistic particles. This mechanism bridges the kHz waves induced by crustal oscillations and the observed FRB GHz waves through inverse Compton boost by a factor of $\gamma^2$, where $\gamma \sim$ a few hundreds is the Lorentz factor of the FRBs. \cite{Zhang22} argued that this mechanism has a list of merits in interpreting the FRB phenomenology, including the reduced requirement for the degree of coherence due to the much larger emission power of the ICS mechanism and the ability of accounting for polarization properties and narrow spectra of FRBs. However, most of the arguments were presented qualitatively in that paper.

In this paper, we revisit the ICS model for FRBs and investigate several relevant physical processes in much greater detail. The main components of the scenario we explore in this work are described in Fig. \ref{fig:cartoon}. We first consider the generation and propagation of low-frequency fast magnetosonic waves and Alfv\'en waves in the inner magnetosphere and show that the amplitude of fast magnetosonic waves can be much stronger than that assumed in \cite{Zhang22} thanks to Alfv\'en waves - fast magnetosonic waves conversion. On the other hand, we point out that the ICS emission power of individual particles is significantly lower than estimated by \cite{Zhang22}, who adopted an O-mode-to-O-mode scattering cross section that is valid only in vacuum. The combination of these two corrections still ensures that the ICS emission power is much greater than the CR power at emission radii beyond $\sim 100 R_\star$, where $R_\star$ is the neutron star radius. Another important ingredient unveiled in this study is that we demonstrate that the low-frequency fast magnetosonic waves not only serve as the seed photons for relativistic particles to upscatter, but also serve as an agent to naturally redistribute relativistic particles into bunches with the size of the FRB emission wavelength in the observer frame. As a result, the bunch formation and maintenance problems are naturally solved.  We also demonstrate that the ICS model prediction can satisfy many observational constraints of FRB emission, including narrow spectra and polarization properties, and hence, should be seriously considered as a strong candidate of the magnetospheric radiation mechanism of FRBs. 
{The model is specifically developed to address observations of repeating FRBs, but the model may apply to  apparently non-repeating FRBs with a magnetar origin, even though we do not exclude the possibility that other engines/mechanisms might also at play to interpret some apparently non-repeating FRBs.}

This paper is organized as follows. In Section \ref{sec:low frequency wave}, we investigate the physics of low frequency waves in a magnetized plasma and delineate the properties of fast magnetosonic waves in a magnetar magnetosphere. In Section \ref{sec:ICS model}, we introduce the basic properties of coherent magnetospheric ICS model and in Section \ref{sec:confront} we confront the model predictions (spectral and polarization properties) against the observational results. A comparison between the coherent ICS and CR models are discussed in Section \ref{sec:comparison}. Conclusions are summarized in Section \ref{sec:conclusion}.

\section{Low-frequency waves in a magnetized plasma}\label{sec:low frequency wave}

Starquakes and crustal crackings are commonly invoked to trigger FRB emission from a magnetar. These events are accompanied with crustal oscillations, which are naturally accompanied by low-frequency waves, typically with frequencies in the range of $10^3 - 10^4$ Hz. These waves can propagate through the crust with a height $h \lesssim 0.1R_\star\sim(10^5 \ {\rm cm}) \ R_{\star,6}$) and with a speed $v\sim 0.01c$. The period of magnetosonic waves can be roughly estimated as $\Delta t_{\rm fmw} \gtrsim h/v \sim (1 \ {\rm ms}) \ R_{\star,6}$, thus the characteristic frequency of the waves is estimated as $\omega_i\sim10^4$ Hz \citep{Blaes1989,Bransgrove2020}.
In the magnetar magnetospheres with a strong magnetic field,  the electromagnetic waves have two orthogonal modes when the wave frequency is larger than the plasma frequency in the rest frame of the plasma ($\omega' > \omega'_p$, where $\omega'_p$ is the plasma frequency in the plasma co-moving frame), i.e. the X-mode whose polarization is perpendicular to the $\vec k-\vec B_{0}$ plane and the O-mode whose polarization is parallel to the $\vec k-\vec B_{0}$ plane. 
The low-frequency waves studied in this paper, on the other hand, are in the $\omega' \ll\omega'_p$ regime. In this regime, the two modes are the Alf\'ven wave mode with polarization parallel to the $\vec k-\vec B_{0}$ plane, which can only propagate along background magnetic field, and the fast magnetosonic wave mode with polarization perpendicular to the $\vec k-\vec B_{0}$ plane, which can propagate freely and share identical properties as the X-mode. At an emission radius $r$, the plasma frequency in the comoving frame can be estimated as
\begin{equation}
\begin{aligned}
\omega_p'&=\sqrt{\frac{4\pi e^2\xi n_{\rm GJ}'}{ m_e}}\simeq\sqrt{\frac{4\pi e\xi B_\star}{\gamma Pm_ec}}\left(\frac{r}{R_\star}\right)^{-3/2}\\
&\simeq(4.7\times10^7 \ {\rm rad \ s^{-1}}) \ \xi^{1/2}B_{\star,15}^{1/2}P^{-1/2} r_8^{-3/2}R_{\star,6}^{3/2}\gamma_{2}^{-1/2},
\end{aligned}
\end{equation}
which is much greater than the kHz waves generated from the magnetar surface in the comoving frame,
i.e. $\omega_i'\sim\gamma \omega_i=(10^6 \ {\rm Hz}) \ \gamma_2\omega_{i,4}$ for both fast magnetosonic waves and Alfv\'en waves. 
Here $\xi$ is the pair multiplicity factor with respect to the Goldreich Julian density $n_{\rm GJ}$, which is normalized to unity here, $P$ is the magnetar period normalized to unity, $\gamma$ is the bulk motion Lorentz factor of the emitting particles, 
and $B_{\star}$ is the magnetic field strength at the magnetar surface. The cyclotron frequency is
\begin{equation}
\omega_B\simeq(1.8\times10^{16} \ {\rm rad \ s^{-1}}) \ B_{\star,15} r_8^{-3}R_{\star,6}^3
\end{equation}
implying that the plasma particles are  constrained to the lowest Landau energy level.

In the following, we discuss the basic properties of the two modes of low frequency waves and their propagation in a magnetized pair plasma relevant to a magnetar magnetosphere.
We consider that the background magnetic field $\vec B_0$ is along $z$-axis and that the wave vectors for both Alf\'ven waves and fast magnetosonic waves ($\vec k_{\rm aw}$ and $\vec k_{\rm fmw}$) lie in the $x-z$ plane with an angle $\theta$ with respect to the $z$-axis.

\subsection{Alf\'ven waves and fast magnetosonic waves}
In this subsection, we briefly discuss the dispersion relations of Alf\'ven waves and fast magnetosonic waves. 

The dispersion relation for Alf\'ven waves (see Appendix \ref{App:dispersion relation} for a derivation) is
\begin{equation}
\omega^2=k^2c^2\cos^2\theta,
\end{equation}
where $\theta$ is the angle between the wave vector and the background magnetic field.
In the left panel of Fig. \ref{fig:alfven and fmw}, only $E_x$ can exist for Alf\'ven waves (see Appendix \ref{App:dispersion relation} for a detailed derivation).
One can see that the perturbation to the background magnetic field is $B_{\rm aw}\hat{y}$ determined by $\vec k_{\rm aw}\times\vec E_{\rm aw}$. Then the group velocity and Poynting flux have the same direction along $z$-axis. $\vec E_{\rm aw}$ is parallel to the $\vec k_{\rm aw}-\vec B_0$ plane, and Alf\'ven waves can only propagate along $\vec B_0$.
The force-free MHD equation $\vec E+(\vec v\times\vec B)/c=0$ gives the velocity $\vec v=-(E_x/B_0)\hat{y}$ (green arrow line in the left panel of Fig. \ref{fig:alfven and fmw}). 
The background magnetic field $\vec B_{0}$ is sheared by $\vec B_{\rm aw}$.

\begin{figure}
\includegraphics[width=\columnwidth]{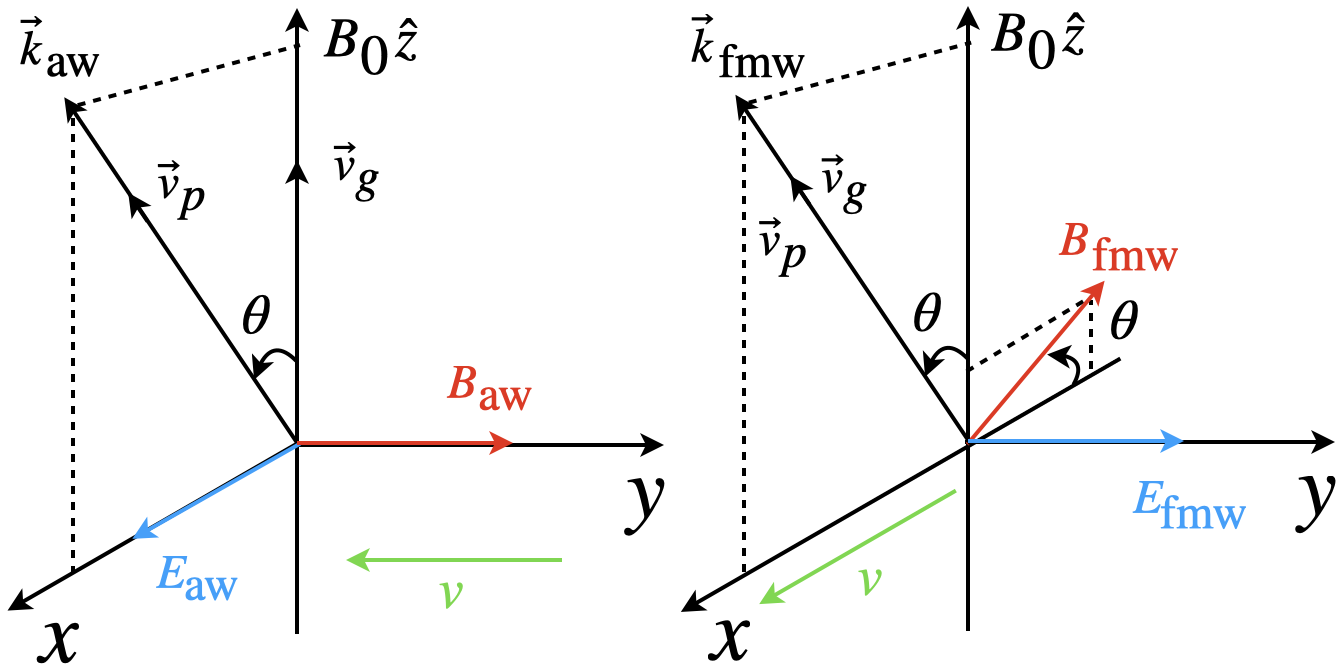}
    \caption{Relative directions for the waves vectors, electric (blue arrow line) and magnetic (red arrow line) fields, phase and group velocity and velocity of plasma (green arrow line) for the Alf\'ven waves (left panel) and fast magnetosonic waves (right panel). The background magnetic field $\vec B_0$ is assumed to be along $z$-axis and the wave vector is in the $x-z$ plane.}
    \label{fig:alfven and fmw}
\end{figure}

The dispersion relation for fast magnetosonic waves is (see Appendix \ref{App:dispersion relation} for a derivation)
\begin{equation}
\omega^2=k^2c^2.
\end{equation}
In the right panel of Fig. \ref{fig:alfven and fmw}, $\vec E_{\rm fmw}$ is along $y$-axis ($E_x=0$ in this case, see Appendix \ref{App:dispersion relation} for a detailed derivation) and perpendicular to the $\vec k_{\rm fmw}-\vec B_0$ plane. The wave magnetic field is determined by $\vec k_{\rm fmw}\times\vec E_{\rm fmw}$.
The phase velocity and group velocity have the same direction along the wave vector. The force-free condition gives the velocity $\vec v=(E_y/B_0)\hat{x}$ (green arrow line in the right panel of Fig. \ref{fig:alfven and fmw}). One can see that the wave vector has a non-zero component along $\vec v$, when the waves propagate in the direction perpendicular to $\vec B_0$, $\vec B_{\rm fmw}$ is parallel to $\vec B_0$, i.e. the background magnetic field lines are compressed and not sheared.

\subsection{Propagation of fast magnetosonic waves inside the magnetosphere and inverse Compton scattering}\label{sec:propagation of LF wave}

In this subsection, we discuss the propagation properties of low-frequency waves, including the strong wave effect of magnetosonic waves and polarization states of the two modes.

We assume that in large scale the magnetic field configuration of the magnetar is dipolar. The background magnetic field therefore scales with $B_0\propto r^{-3}$. The low-frequency waves have different scaling relations. To ensure energy conservation, the amplitude of the Alf\'ven waves scales as $B_{\rm aw} \propto r^{-3/2}$ and that of fast magnetosonic waves scales as $B_{\rm fmw}\propto r^{-1}$ due to spherical expansion. 
Since the amplitudes of both low-frequency waves decay more slowly than the background $B_0$, they could become greater than $B_0$ at a large enough radius, where non-linear wave effects may play an important role.
The decay of fast magnetosonic waves into Alf\'ven waves could also happen in the non-linear regime \citep{Golbraikh&Lyubarsky2023}. On the other hand, the reverse process,  i.e. the conversion of Alf\'ven waves into fast magnetosonic waves, is also possible, as is investigated in the closed filed line region for small-amplitude Alf\'ven waves \citep{Yuan2021}. For strong fast magnetosonic waves, which may not suffer much from dissipation, perpendicular shocks could be formed in the closed field line region \citep{ChenYR2022}.
{It should be pointed out that only the component of fast magnetosonic waves with a magnetic field in the direction opposite to the background magnetic field lines can develop the non-linear regime.}
However, for the coherent ICS model studied in this paper, the required low-frequency waves would not enter the non-linear regime. 
{This is because in the open field line region, it is hard to have the total electric field strength to be above the total magnetic field strength.}
Avoiding the non-linear regime is also needed to account for the observed narrow frequency spectra of FRBs (see Sect.\ref{sec:spectra}).

\begin{figure}
\includegraphics[width=95mm]{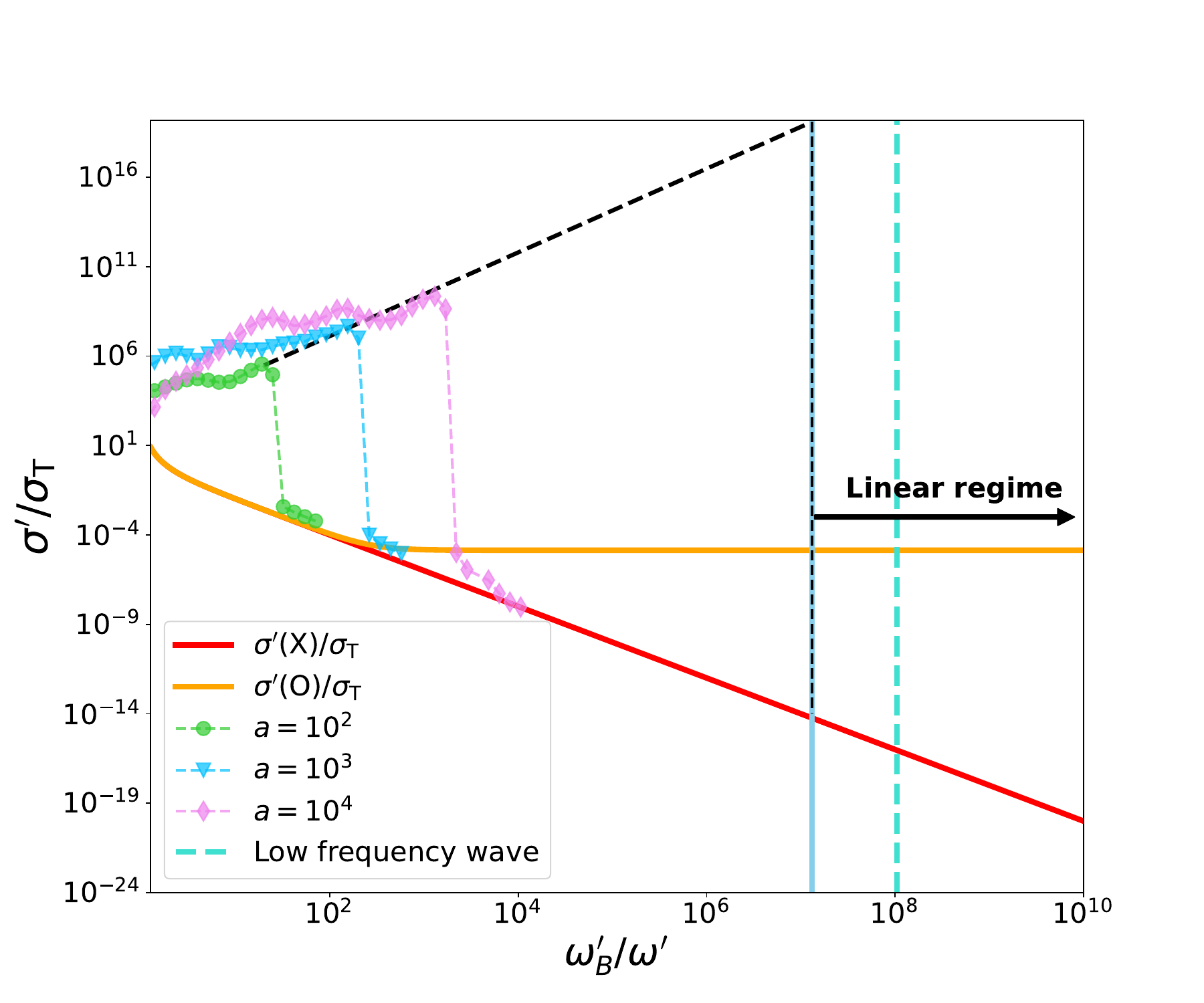}
    \caption{The normalized cross section ($\sigma'/\sigma_{\rm T}$) for X-mode (red line) and O-mode (orange line) in the linear regime, and for different large non-linear parameters ($a\gg1$) waves as a function of $\omega_B'/\omega'$. 
    The following parameters are adopted: the lepton Lorentz factor $\gamma=800$, incident angle of the low frequency wave $\theta_i=30^\circ$, magnetar surface magnetic field strength $B_\star=10^{15}$ G, magnetar radius $R_\star=10^6$ cm, incident waves frequency $\omega_i=10^4$ Hz, and initial wave amplitude at the magnetar surface $E_{\rm fmw,0}=10^{9}$ esu.}
\label{fig:no nonlinear}
\end{figure}

Let us consider the interaction between the incident fast magnetosonic waves and relativistic plasma particles inside the magnetar magnetosphere.  
A commonly invoked parameter is the amplitude parameter
\begin{equation}
a=\frac{qE_w}{m_ec\omega_i},
\end{equation}
where $E_w$ and $\omega_i$ are the electric field amplitude and frequency of the incident low-frequency waves, respectively. 
This parameter is Lorentz invariant, keeping the same form in both the co-moving frame of the relativistic particles and the lab frame 
(see Appendix \ref{App:transformation} for a derivation of the waves' electric field transformation), i.e. 
\begin{equation}
a'=\frac{qE_{w}'}{m_e c\omega_i'}=\frac{qE_{w}\gamma(1-\beta\cos\theta_i)}{m_e c \omega_i \gamma (1-\beta\cos\theta_i)}=a,
\end{equation}
with $a'$ being relevant for considering the inverse Compton scattering process. One can estimate the amplitude parameter of the incident low-frequency waves as 
\begin{equation}\label{eq:a-parameter for incident wave}
\begin{aligned}
a&=\frac{qE_{\rm w,0}}{m_ec\omega_i}\left(\frac{r}{R_\star}\right)^{-1}\simeq3.5\times10^{9} \ E_{w,0,9}\omega_{i,4}^{-1}r_{8.7}^{-1}R_{\star,6},
\end{aligned}
\end{equation}
where $E_{\rm w,0}$ is the initial amplitude of the incident waves at the magnetar surface, which is normalized to $10^{9}$ as required by the model (see more details below). We fix the Lorentz factor to $\gamma=800$ and present the cross section for wave-particle interaction normalized to the Thomson cross section ($\sigma'/\sigma_{\rm T}$) as a function of $\omega_B'/\omega'$  in the left panel Fig. \ref{fig:no nonlinear}. Solid lines are the  X-mode (red line) and the O-mode (orange line) cross sections in the linear regime (see Eq.(\ref{eq:Xmode}) below and Appendix \ref{App:cross section} for a detailed derivation of the cross sections). We also present the numerical data of the X-mode $\sigma'/\sigma_{\rm T}$ as a function of $\omega_B'/\omega'$ for large-amplitude waves with different amplitude parameters, i.e. $a=10^2$ (green curves), $a=10^3$ (blue curves) and $a=10^4$ (purple curves) for the case of $\theta_i'=0.1$, using the same data as presented in the upper panel of Fig.3a in \cite{QKZ}. One can see that there is a clear transition at \citep{QKZ}\footnote{The physical understanding of this transition is as follows: The amplitude parameter $a$ can be also written as $\omega_{B_w}/\omega$, where $\omega_{B_w} = eB_w/m_ec$. The large amplitude effect kicks in when the $B_w$ component along the background $B_0$ becomes comparable to $B_0$ itself, i.e. $a \sin \theta \sim \omega_B / \omega$. When $\theta$ is small, one has $\sin\theta \sim \theta$ and we get Eq.(\ref{eq:linear}).}
\begin{equation}
\frac{\omega_B'}{\omega'}\simeq a'\theta_i',
\label{eq:linear}
\end{equation}
above which the scattering enters the linear regime and the cross section significantly falls to be consistent with the analytical linear approximation. This suggests that the scattering in the linear regime even if $a' \gg 1$ as long as the scattering happens in the strongly magnetized regime. We also tested the numerical data for $\theta_i'=10^{-2}$ and $\theta_i'=10^{-3}$ for the three $a$ values calculated in \cite{QKZ} and obtained the same trend. Hereafter, we adopt Eq.(\ref{eq:linear}) as the criterion to define the linear-non-linear regime transition, even if the $a$ value of low-frequency waves is much higher than that numerically explored in \cite{QKZ}.

We fix the incident angle $\theta_i=30^{\circ}$ in the lab frame. We  present the scaling line (black dashed) corresponding to the peak of the cross section when $a$ increases until the value of Eq.(\ref{eq:a-parameter for incident wave}) is reached. One can see that the peak value is proportional to $\sim a^2$ and drops significantly after the transition point. The scattering is in the linear regime if 
\begin{equation}
\frac{\omega_B'}{\omega'} > a'\theta_i'\simeq1.6\times10^7  E_{w,0,9}\omega_{i,4}^{-1}r_{8.7}^{-1}R_{\star,6}
\end{equation}
is satisfied. We present the blue vertical solid line corresponding to the value of $a'\theta_i'$ in Fig.\ref{fig:no nonlinear}.
For $\omega_i=10^4 \ {\rm rad \ s^{-1}}$, one has $\omega_B'/\omega_i'\simeq4.2\times10^{10}$ at $r \sim500R_\star$ (dashed turquoise line in Fig. \ref{fig:no nonlinear}). Therefore, the low frequency waves would be in the linear regime for the ICS process we propose in this paper. The non-linear regime of cross section also warrants narrow spectra as discussed in Sect. \ref{sec:spectra}.

\subsection{Origins of fast magnetosonic waves}

One way of generating fast magnetosonic waves is through conversion from Alf\'ven waves \citep{Yuan2021,Chen2024,Mahlmann2024}. The conversion is usually described by three-wave interactions, with an Alf\'ven wave with wave number vector $\vec k_1$ interacting with a virtual wave defined by the background magnetic field with wave number vector $\vec k_2$ whose amplitude is much smaller than that of the Alf\'ven waves, i.e. $k_2 \ll k_1$. Spatially isotropic fast magnetosonic waves with $\vec k$ can be generated when the energy conservation law is satisfied, i.e. $\vec k_1+\vec k_2=\vec k$. If one begins with pure Alf\'ven waves, the energy conversion efficiency is proportional to $(r/R_{\rm LC})^2$ \citep{Yuan2021} (also see Appendix \ref{App:dispersion relation} for a detailed derivation). It should be pointed out that the conversion efficiency is mainly related to the radius from surface normalized to the light cylinder, i.e.
\begin{equation}
f_{\rm A\rightarrow F}=\frac{E_{\rm fmw}^2}{E_{\rm aw}^2}\simeq\left(\frac{r}{R_{\rm LC}}\right)^2\simeq0.01 \ r_{8.7}^2P^{-2},
\end{equation}
where $r=500R_{\star}$ is applied for the typical radiation radius for the ICS model. Notice that the conversion efficiency is 
independent of the wave frequency.
One can see that $E_{\rm fmw}\simeq0.1 E_{\rm aw}$ at the relevant radius. Even if magnetar crust cracking only generates Alf\'ven waves with a relative amplitude $B_{\rm aw,0}/B_0\sim10^{-4}$, the fast magnetosonic waves can still have a large enough amplitude to power FRBs through the conversion process.

Second, fast magnetosonic waves may be directly generated from the surface region due to the horizontal shear perturbation at the crust. Indeed, such a shearing process can excite a mixture of Alf\'ven and magnetosonic waves as long as the magnetic field is not exactly perpendicular to the surface \citep{Yuan2022}. A detailed study of this process will be postponed in future work. In the following, we do not specify the mechanism of fast magnetosonic wave generation but adopt an effective amplitude at surface if they were produced from the surface region.

\section{Coherent inverse Compton scattering by bunches}\label{sec:ICS model}

\subsection{Emission frequency}

The ICS process can boost the kHz waves to the observed GHz FRB waves. When the incident fast magnetosonic waves reach the typical emission radius, the bunched particles would oscillate at the same frequency in the comoving frame, so the frequency of the scattered waves through ICS in the lab frame can be written as
\begin{equation}
\nu=\gamma^2\nu_i(1+\beta\cos\theta_1')(1-\beta\cos\theta_i)\sim \gamma^2\nu_i\simeq (1 \ {\rm GHz}) \ \gamma_{2.9}^2\omega_{i,4},
\end{equation}
where $\theta_i$ is the incident angle in the lab frame and $\theta_1'$ is the scattered angle in the comoving frame, and the particle Lorentz factor $\gamma$ is normalized to $800$ to produce 1-GHz waves.
The boosting factor $\gamma^2$ is valid for both non-magnetized and magnetized cases, as long as the charged particles have non-relativistic motion in the comoving frame.

\subsection{Scattering power of a single charged particle}\label{sec:emission power}

In the comoving frame (along $B_0\hat{z}$) of a relativistic lepton (electron or positron), the Compton scattering cross section in a strong magnetic field for X-mode waves is given by (see Appendix \ref{App:cross section} for details) \citep{Herold1979}
\begin{equation}
\sigma'({\rm X})=\frac{\sigma_{\rm T}}{2}\left[\frac{\omega'^2}{(\omega'+\omega_B)^2}+\frac{\omega'^2}{(\omega'-\omega_B)^2}\right],
\label{eq:Xmode}
\end{equation}
where $\omega_B'=\omega_B$ is applied. The fast magnetosonic waves are X-mode waves and can propagate freely without much scattering in the strong magnetic field region since the cross section is greatly suppressed with $\sim\sigma_{\rm T}(\omega'/\omega_B)^2\ll\sigma_{\rm T}$.
In the lab frame, the cross section is related to the comoving frame Compton scattering cross section through the transformation $\sigma=(1-\beta\cos\theta_i)\sigma'$ and we have
\begin{equation}
\sigma({\rm X})=\frac{\sigma_{\rm T}}{2}(1-\beta\cos\theta_i)\left[\frac{\omega'^2}{(\omega'+\omega_B)^2}+\frac{\omega'^2}{(\omega'-\omega_B)^2}\right],
\end{equation}
where $\omega'=\gamma(1-\beta\cos\theta_i)\omega\sim\gamma\omega$ for $\theta_i>0$.
{The X-mode cross section is obtained by considering one single particle's radiation in the incident X-mode waves and a uniform background magnetic field. For a plasma, the cross section of the collective Thomson scattering in strong magnetic fields is essentially the same as that of a single particle \citep{Nishiura&Ioka2024}.}

The photon energy density of the incident low frequency waves at an emission radius $\sim500R_\star$ can be estimated as
\begin{equation}
U_{\rm ph}\simeq\frac{E_{\rm fmw,0}^2}{4\pi}\left(\frac{r}{R_\star}\right)^{-2}\simeq (3.2\times10^{11} \ {\rm erg \ cm^{-3}}) \  E_{\rm fmw,0,9}^2r_{8.7}^{-2}R_{\star,6}^2.
\end{equation} 
Then the ICS emission power of a single relativistic lepton can be calculated as
\begin{equation}\label{eq:ICS single power}
\begin{aligned}
P_{\rm ICS}&
\simeq\gamma^2\sigma({\rm X})cU_{\rm ph}\\
&\simeq(3.2\times10^{-14} \ {\rm erg \ s^{-1}}) \ \gamma_{2.9}^4\omega_{i,4}^2B_{\star,15}^{-2} E_{\rm fmw,0,9}^2r_{8.7}^{4}R_{\star,6}^{-4},
\end{aligned}
\end{equation}
which depends on the particle Lorentz factor (normalized to $\sim 800$ required to scatter $\omega_i=10^4 \omega_{i,4}$ waves to the GHz wave), the cross section 
and the incident wave energy density.
One can compare this power with that of curvature radiation. In order to produce 1 GHz waves via curvature radiation, one requires $\gamma_{\rm CR}\simeq412(\rho_{8.7}\nu_9)^{1/3}$, $\rho$ is the curvature radius of the magnetic field line. Then the emission power of CR can be calculated as $P_{\rm CR}=2e^2c\gamma_{\rm CR}^4/(3\rho^2)\simeq5.3\times10^{-16} \ {\rm erg \ s^{-1}} \ \nu_9^{4/3}\rho_{8.7}^{-2/3}$. 
From Eq.(\ref{eq:ICS single power}), one can see that $P_{\rm ICS}\gg P_{\rm CR}$ at the normalized radius.

\begin{figure}
\includegraphics[width=95mm]{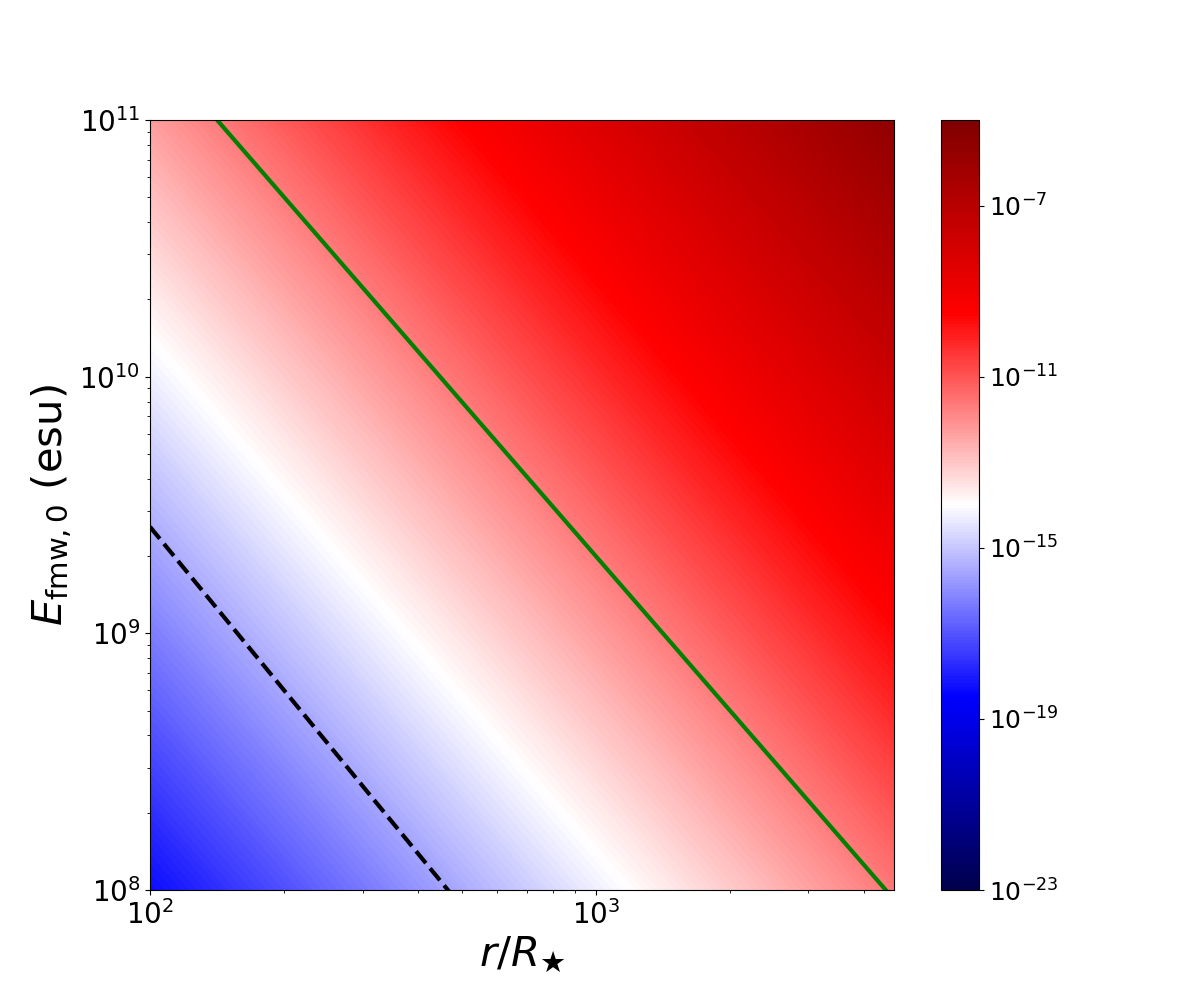}
    \caption{The ICS emission power of individual particle as a function of normalized radius $r/R_\star$ and initial amplitude of fast magnetosonic waves $E_{\rm fmw,0}$ at the magnetar surface. Following parameters are adopted: magnetar surface magnetic field strength $B_\star=10^{15}$ G, magnetar radius $R_\star=10^6$ cm, Lorentz factor of leptons $\gamma=800$, incident waves frequency $\omega_i=10^4$ Hz and magnetar period $P=1$ s. The dashed black line marks the typical curvature radiation power of one single particle to produce 1-GHz waves.}
\label{fig:emission power contour}
\end{figure}

\begin{figure}
\includegraphics[width=95mm]{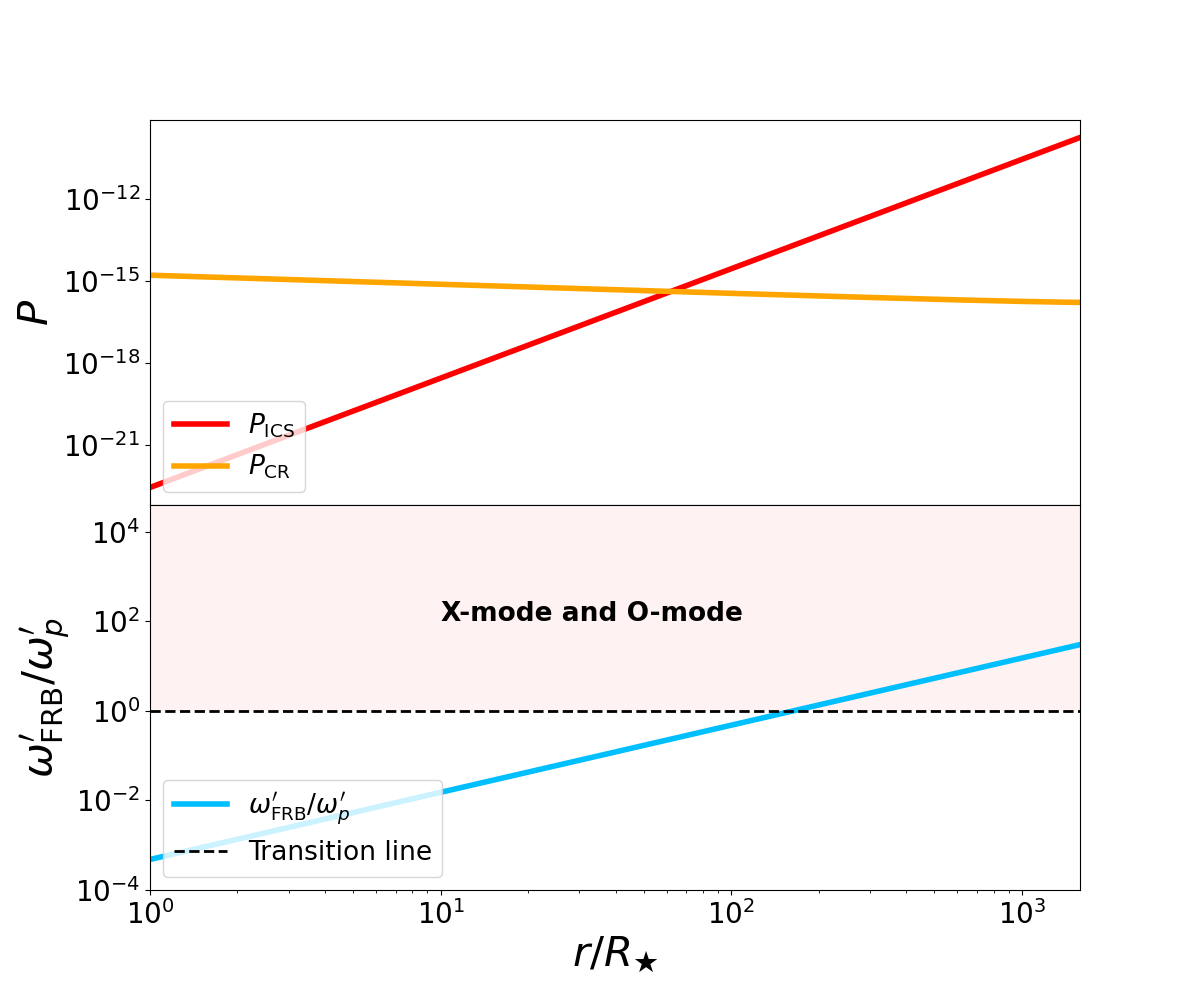}
    \caption{The ICS emission power (red solid line) and CR power (orange solid line) of an individual particle as a function of normalized emission radius $r/R_\star$ in the upper panel. 
    The ratio of typical FRB frequency to plasma frequency ($\omega_{\rm FRB}'/\omega_p'$) in the comoving frame as a function of normalized emission radius in the lower panel. 
    The following parameters are adopted: the Lorentz factor of leptons $\gamma=800$, incident angle of the low frequency wave $\theta_i=30^\circ$, magnetar surface magnetic field strength $B_\star=10^{15}$ G, magnetar radius $R_\star=10^6$ cm, magnetar period $P=1$ s, incident waves frequency $\omega_i=10^4$ Hz, and initial wave amplitude at the magnetar surface $E_{\rm fmw,0}=10^{9}$ esu. The pair multiplicity $\xi$ is normalized to unity, so that the plasma frequency is defined by the Goldreich Julian density.}
    \label{fig:ratio function}
\end{figure}

In Fig. \ref{fig:emission power contour}, we present the ICS emission power of an individual particle as a function of normalized radius $r/R_\star$ and initial amplitude of fast magnetosonic waves $E_{\rm fmw,0}$.
We also present the curvature radiation power for 1-GHz waves as a function of  emission radius (black dashed line).
One can see that the ICS power is larger than the CR for the parameter region above the dashed black line\footnote{It should be pointed out that curvature radiation is more efficient than ICS when the emission region is close to the magnetar surface. However, the magnetospheric model of FRBs prefers a large emission radius for the following three reasons: (i) FRBs produced near surface suffers a higher scattering optical depth before escaping the magnetosphere; (ii) A strong $E_\parallel$ required to sustain FRB emission more likely survives at a larger emission radius because the dense electron-positron pairs envisaged in the near-surface region tend to screen any $E_\parallel$ there ($E_\parallel$); (iii) At lower emission radii, the GHz waves tend to be below the plasma frequency so that only X-mode waves can escape. Emission from these radii would not have circular polarization (see Fig.\ref{fig:ratio function}). }. The green solid line denotes the transition from the linear to the non-linear regime, i.e. the linear regime is below the green line. 
Therefore, the linear ICS process will occur between the green solid line and the black dashed line.

Another way to compare the emission power of the two emission mechanisms is through comparing the single particle's emission power of ICS and CR as a function of the normalized emission radius, as shown in the upper panel of Fig. \ref{fig:ratio function}.
One can see that for the nominal value of $E_{\rm fmw,0} =10^9$ esu, ICS dominates over CR at $r\gtrsim100R_\star$. 
In the lower panel of \ref{fig:ratio function}, we calculate $\omega_{\rm FRB}'/\omega_p'$ as a function of normalized radius. The red region above the transition line ($\omega_{\rm FRB}'=\omega_p'$) denotes the region that both X-mode and O-mode can propagate freely. One can see that this also happens at $\gtrsim100R_\star$.
We take the magnetar period $P=1$ s to calculate the blue solid line and the light cylinder radius (green dashed line). 
It should be pointed out that the incident low frequency waves are in the fast magnetosonic mode or X-mode, but the scattered waves could have both X-mode and O-mode. This will allow the ICS mechanism to produce a variety of FRB polarization properties (see  Sect.\ref{sec:polarization}).

\subsection{Bunch formation due to the low-frequency magnetosonic waves} \label{sec:bunching}

Different from the curvature radiation model that requires a separate physical mechanism (e.g. two stream plasma instabilities) to generate bunches, the low-frequency fast magnetosonic waves can naturally redistribute relativistic leptons in the emission region, forming bunches with the right size for FRB emission (see \citealt{Zhang2023Universe} for a brief discussion). By definition, the incident electric field of fast magnetosonic waves is perpendicular to the background magnetic field (along $z$-axis), and in the comoving frame it can be written as $\vec E_{\rm fmw}'=E_{\rm fmw}'\cos(\omega_i' t')\hat{x}$.
The velocity of a charged particle with charge $q$ in the wave electric field $\vec E_{\rm fmw}'$ and background magnetic field $B_0\hat{z}$ can be calculated as\footnote{For simplicity, we ignore $\vec B_{\rm fmw}'$ with the assumption that it is small compared with the background $B_0\hat{z}$. This is generally valid for the parameter space we discuss in this paper. If $\vec B_{\rm fmw}'$ becomes large enough, the particle trajectories become more complicated, and the simple harmonic treatment here should be modified. }
\begin{equation}\label{eq:bunch formation velocity}
\vec v'=\frac{qE_{\rm fmw}'}{m_e(\omega_i'^2-\omega_B'^2)}[\omega_i'\sin(\omega_i' t')\hat{x}+{\omega_B'}\cos(\omega_i't')\hat{y}].
\end{equation}
The corresponding displacement can be integrated as
\begin{equation}\label{eq:bunch formation}
\begin{aligned}
\vec r'&=\frac{qE_{\rm fmw}'}{m_e(\omega_i'^2-\omega_B'^2)}\left[\frac{\omega_B'}{\omega_i'}\sin(\omega_i't')\hat{y}-\cos(\omega_i' t')\hat{x}\right]\\
&\simeq\frac{a'c\omega_B'}{\omega_i'^2-\omega_B'^2}\sin(\omega_i't')\hat{y}.
\end{aligned}
\end{equation}
From Eq.(\ref{eq:bunch formation velocity}), one can see that there are two harmonic oscillation directions: along $x$-axis which is due to the incident wave's electric field and along $y$-axis which is due to the $\vec E_{\rm fmw}'\times\vec B_0$ drift. 
The cross section contributed by the $x$ and $y$ directions are proportional to $(\omega_i'/\omega_B')^4$ and $(\omega_i'/\omega_B')^2$, respectively (see Appendix \ref{App:cross section} for a derivation). So the dominant displacement is in the $y$ direction (drifting direction). Notice that 
the acceleration along $y$-axis is independent of charge so that a non-zero net charge is needed to sustain significant radiation in the $y$-direction, i.e. a significant ICS process requires bunches with a non-zero net charge.

\begin{figure*}
\includegraphics[width=17 cm,height=8.5 cm]{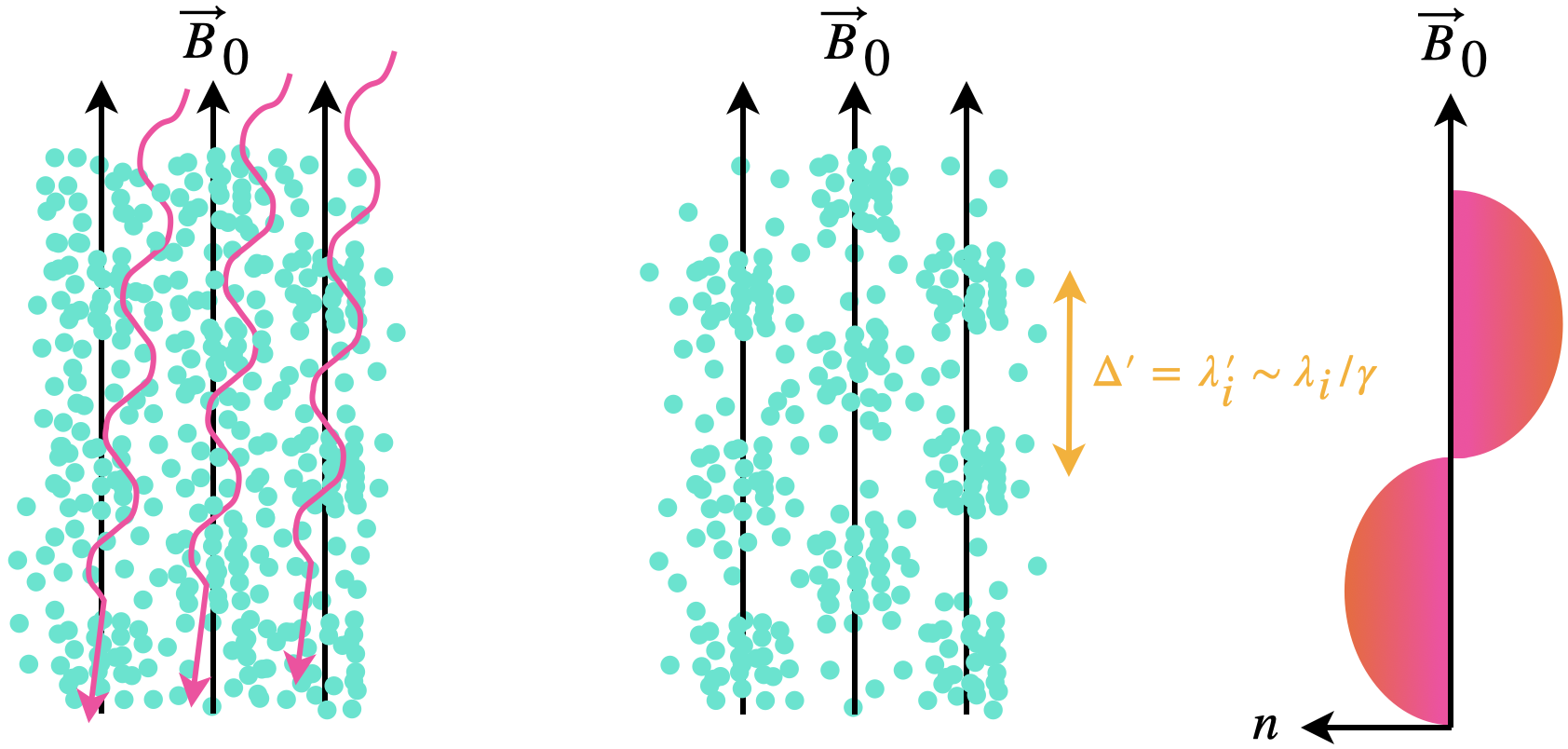}
\caption{This cartoon figure explains a schematic bunch formation mechanism due to the incident low frequency fast magnetosonic waves described in this work. The comoving frame is used, in which the boosted low-frequency waves move along the nearly opposite direction of the particle motion (the $\vec B_0$ direction). The purple wiggler denotes the incident low frequency waves. Green dots denote single-specie particles. Low frequency waves will induce spatial clumps, with the separation between clumps $\Delta=\lambda'\sim\lambda_i/\gamma$.}
\label{fig:bunch formation}
\end{figure*}

From Eq.(\ref{eq:bunch formation}), the maximum amplitude of $\vec r$ can be estimated as 
\begin{equation}
|\vec r'|_{\rm max}\simeq \frac{a'c}{\omega_B'}\simeq(7.5\times10^5 \ {\rm cm}) \ E_{\rm fmw,0,9}r_{8.7}^{2}R_{\star,6}^{-2}\omega_{i,4}^{-1}B_{\star,15}^{-1},
\end{equation}
which means that the lepton will be pulled away from the initial position by $\sim (0-10^6)$ cm due to the incident fast magnetosonic waves with a density fluctuation profile in a rough sinusoidal form.
The $x$-component can be ignored since the amplitude is smaller than the $y$-component by a factor of $\omega_B'/\omega_i'$. Therefore, when the incident fast magnetosonic waves propagate with the angle $\theta_i$ with respect to $\vec B_0$, the wave electric field along $x$-axis would induce a drift notion in $y$-axis, i.e. the particles would be separated and collected sinusoidally and charged bunches would be naturally produced. 
In the comoving frame of the relativistic lepton, the photon incident angle can be calculated as
\begin{equation}
\sin\theta_i'={\cal D}\sin\theta_i \ \Rightarrow \ \theta_i'\simeq4.7\times10^{-3} \ {\rm rad}
\end{equation}
for $\gamma=800$ and $\theta_i=30^\circ$, i.e. the incident waves nearly propagate along the background magnetic field lines in the opposite direction of the particle's moving direction in the comoving frame, i.e. 
$\theta_i'$ is defined as the angle between the opposite direction of the incident wave vector and $\hat{z}$ in the comoving frame. The wavelength of the incident waves is de-boosted by $\sim\gamma$ in the comoving frame and the particle spatial distribution would follow a pattern with a typical length $\sim\lambda_i/\gamma$. This length in the lab frame is $\sim\lambda_{i}/\gamma^2\sim\lambda_{\rm frb}$, i.e. the longitudinal size of the charged bunch has the same size as the FRB wavelength.
Thus, the scattered waves can be considered as coherent since the longitudinal size of the bunch is smaller than $\lambda_{\rm frb}$.

We present a cartoon figure to explain the bunch formation mechanism in the comoving frame in Fig. \ref{fig:bunch formation}: In the left panel, a single-specie plasma is nearly homogeneously (but with small fluctuations) distributed in the typical emission region before interacting with the fast magnetosonic waves. In the central panel, the charged bunches are produced by the incident fast magnetosonic waves and the separation is $\Delta'=\lambda'\sim\lambda_i/\gamma$. In the right panel, the number density of the plasma along $\vec B_0$ is displayed.

In the $y$-axis perpendicular to $z$-axis, i.e. the direction of $\vec E_{\rm fmw}\times\vec B_0$ drift, the number density would fluctuate sinusoidally. This would induce a sinusoidal fluctuation along $z$-axis with the coherence factor of one single bunch's radiation described as 
\begin{equation}
\left|\sum_j^{N_{e,b}} {\rm exp}\left[-i\omega\left(\frac{\hat{n}\cdot{\Delta\vec r_j}}{c}\right)\right]\right|^2,
\end{equation}
where $N_{e,b}$ is the total number of net charges in one bunch, $\hat{n}$ is the unit vector along the line of sight and $\Delta r_j$ denotes the relative separation distance between the first particle and the $j$th particle in the bunch. 
It should be pointed out that the radiation is mainly within the $\sim 1/\gamma$ cone centered along $z$-axis and the direction of charge separation is perpendicular to $\vec B_0$. Hence, $\hat{n}\cdot\Delta\vec r_j\simeq0$ and the coherent factor is $\sim N_{e,b}^2$, i.e. the transverse displacement due to the $\vec E_{\rm fmw}'\times\vec B_0$ drift would not influence the coherent condition.

We then consider $N_b$ radiation regions that are radiating photons with electric field $E_j(z-z_j)$ and perform Fourier transformation to obtain the power spectrum which is proportional to $E^2(k)$, i.e. \citep{Rybicki&Lightman1979,Yang2023}
\begin{equation}
I(k_s)\propto E^2(k_s)=\left|\sum_j^{N_b}  E_{0,j}^2(k_s)e^{i k_s z_j}\right|^2, 
\end{equation} 
where $N_b$ is the total number of the bunches, $k_s$ is the scattering wave number and $E_{0,j}$ denotes the electric field amplitude of the $j$-th charged bunch's radiation.
Then the coherence factor of the $N_b$ emitting bunches can be calculated as
\begin{equation}\label{eq:coherent factor}
\left|\sum_j^{N_b}  e^{i k_s z_j}\right|^2=\frac{\sin^2(N_bk_s \Delta z/2)}{\sin^2(k_s \Delta z/2)}.
\end{equation}
Therefore, the total emitted FRB luminosity should be calculated as
\begin{equation}\label{eq:coherent luminosity}
\begin{aligned}
L_{\rm frb}&=\frac{\sin^2(N_bk_sc\Delta t/2)}{\sin^2(k_sc\Delta t/2)}N_{e,b}^2P_{\rm ICS}\\
& \simeq \left\{
\begin{aligned}
&N_{b}N_{e,b}^2P_{\rm ICS}, &&{\rm incoherent \ superposition}, \\
&N_{b}^2N_{e,b}^2P_{\rm ICS}, &&{\rm coherent \ superposition},
\end{aligned}
\right.
\end{aligned}
\end{equation}
where $N_{e,b}$ can be estimated as
\begin{equation}
N_{e,b}=\zeta n_{\rm GJ}\pi(\gamma\lambda_{\rm frb})^2\lambda_{\rm frb}=\frac{\pi\zeta B_{\star}\gamma^2\lambda_{\rm frb}^3}{ qc P}\left(\frac{r}{R_{\star}}\right)^{-3},
\end{equation}
where $\zeta$ is the net charge factor and $n_{\rm GJ}$ is the GJ number density. Note that in our model, $\zeta \leq 1$ (charge depletion) is envisaged (see \S\ref{sec:Eparallel} below). 

In the following, we will quantify the fluctuation of the background number density of net charges due to the incident magnetosonic waves.
From Eq.(\ref{eq:bunch formation velocity}), the drifting velocity of the particle in the comoving frame can be written as
\begin{equation}
\vec v_{\rm drift}'=\frac{qE_{\rm fmw}'}{m_e(\omega_i'^2-\omega_B'^2)}{\omega_B'}\cos(\omega_i't'-k_y'y'-k_z'z')\hat{y}.
\end{equation} 
According to the number density continuity equation, one has
\begin{equation}
\frac{\partial n'}{\partial t'}+\nabla'\cdot(n' \vec v_{\rm drift}')=0 \ \Rightarrow \ \frac{1}{n_{\rm bg}'}\frac{dn'}{dt'}\simeq-\frac{dv_{\rm drift}'}{dy'}.
\end{equation}
Noticing the relations $k_y'=k'\sin\theta_i'$ and $\omega_i'=k'c$ and fixing the incident angle $\theta_i=30^\circ$, one can calculate the maximum value of the relative density fluctuation of net charges as
\begin{equation}
\begin{aligned}
\left|\frac{\Delta n}{n_{\rm bg}}\right|_{\rm max}
&=\frac{a\sin\theta_i'}{\omega_B'/\omega_i'}=\frac{E_{\rm fmw,0}}{B_\star}\left(\frac{r}{R_\star}\right)^2\sin\theta_i\\
&\simeq 0.13 \  E_{\rm fmw,0,9}B_{\star,15}^{-1}r_{8.7}^2R_{\star,6}^{-2}.
\end{aligned}
\end{equation}
One can see that this ratio is the one to define whether the wave-particle interaction enters the non-linear regime, i.e. when $a\sin\theta_i'>\omega_B'/\omega_i'$, the charge density fluctuation is greater than the background density and particles will move relativistically in the comoving frame.

\begin{figure}
\includegraphics[width=95mm]{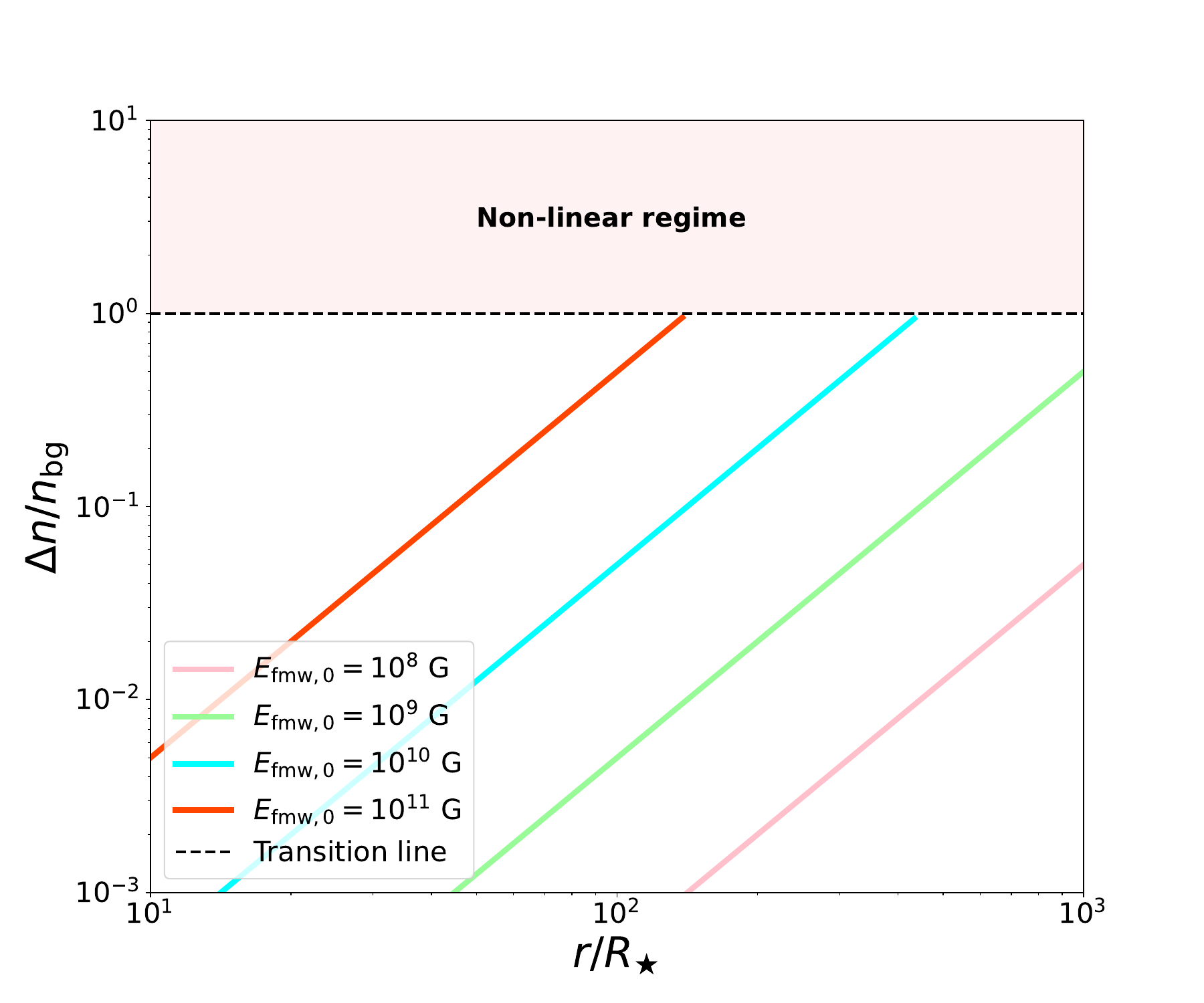}
    \caption{The ratio of total number density fluctuation $\Delta n$ to the background charged number density $n_{\rm bg}$ for different initial fast magnetosonic wave amplitudes as a function of the normalized radius $r/R_\star$. The following parameters are adopted: magnetar surface magnetic field strength $B_\star=10^{15}$ G, magnetar radius $R_\star=10^6$ cm and the incident angle $\theta_i=30^\circ$.}
    \label{fig:fluctuation density}
\end{figure}

We present the ratio of $|\Delta n/n_{\rm bg}|_{\rm max}$ as a function of $r/R_\star$ for different initial wave amplitudes ($ E_{\rm fmw,0}=10^8-10^{11}$ G) in Fig. \ref{fig:fluctuation density}. The non-linear regime ($\Delta n/n_{\rm bg}>1$) is marked as red. A larger $ E_{\rm fmw,0}$ gives particle larger drifting velocity and density fluctuation.

{We note that the above treatment is at the particle level, which delineates a rough physical picture of bunching due to low-frequency fast magnetosonic waves. A more rigorous treatment should be at the plasma level which invokes three-wave interactions in the comoving frame, with an incident wave converted to a plasma density wave and an outgoing X-mode wave. In the lab-frame, the former, likely induced by a ponderomotive force of the low-frequency waves through a higher-order effect, would produce charge bunches with the size of the emission frequency; and the latter would generate the coherent GHz FRB emission. Such an effect is in analogous to the induced Compton scattering effect as discussed by \cite{Nishiura&Ioka2024}, but in strong magnetic fields. The physics of this process deserves a detailed study in a future work. }


\subsection{Charge-depletion, bunch maintenance and the FRB luminosity}\label{sec:Eparallel}

A charge-depleted region within a magnetar's magnetosphere is where a parallel electric field ($E_\parallel$) is developed and particles accelerated. 
For the magnetospheric models for FRBs invoking CR or ICS, an $E_\parallel$ is always needed to continuously supply energy to the rapidly cooling charged bunches in order to reach the high luminosity as observed in FRBs \citep{Kumar2017,Zhang22}. 

The rapid cooling of the bunches requires $E_\parallel$ along the magnetic field lines to continuously provide the high ICS power. The balance between $E_\parallel$ acceleration and ICS radiation cooling requires\footnote{Note that coherent superposition among bunches have been assumed here (Eq.(\ref{eq:coherent luminosity})), which is more likely for low-frequency-wave-induced bunches.}
\begin{equation}
N_{e,b}N_beE_{\parallel} c=N_{e,b}^2N_b^2 \gamma^2\sigma({\rm X})cU_{\rm ph},
\end{equation}
and $E_\parallel$ can be calculated as
\begin{equation}
\begin{aligned}
E_{\parallel}&=N_{e,b}N_b\frac{1}{e}{\gamma^2\sigma({\rm X})U_{\rm ph}}\\
&\simeq (3.6\times10^7 \ {\rm esu}) \ \zeta B_{\star,15}^{-1}\gamma_{2.9}^3P^{-1}\nu_9^{-1}r_{8.7}^{3}R_{\star,6}^{-1}E_{\rm fmw,0,9}^2\\
&\times \omega_{i,4}^2B_{\star,15}^{-2},
\end{aligned}
\end{equation}
where the total number of charged bunches ($N_b$) is defined in Eq.(\ref{eq:total number of bunch}).
Therefore, the relativistic bunches are required to be located in a charge-depleted region. If the bunches exit such a charge depleted region, the bunch would quickly lose energy due to rapid radiative cooling. On the other hand, thanks to its higher power (Section \ref{sec:emission power}), the ICS mechanism allows the background charge density to be sub-Goldreich-Julian in an emission region similar to a ``gap'' envisaged in a pulsar magnetosphere. Once the charged bunches enter the gap region, they will continue to stay in the gap until escaping the magnetosphere.

One natural outcome of having bunches emitting in a charge-depleted region is that the Lorentz factor distribution of the particles is expected to be narrow, i.e. $\Delta\gamma/\gamma_0 \ll 1$.
This is because the leptons with small Lorentz factors ($\gamma<\gamma_0$) would be accelerated by $E_\parallel$ and those with larger Lorentz factors ($\gamma>\gamma_0$) would rapidly cool. As a result, the bunch is essentially mono-energetic, which would result in a narrow spectrum of the scattered waves via the ICS process (see Section \ref{sec:spectra}). The typical Lorentz factor $\gamma \sim 800$ adopted in the ICS model is selected to match the typical observed FRB frequency at GHz. Such a Lorentz factor is achievable thanks to the competition between $E_\parallel$ acceleration and radiative cooling. One may estimate the maximum achievable Lorentz factor of charged particles without radiative cooling, which is $\gamma_{\rm max} \sim q E_\parallel l/(m_ec^2)\sim10^7 \ E_{\parallel,7}l_8$ where $l$ is the typical acceleration length scale of the order of the height of the emission region. The fact that this is much greater than the targeted value $\sim 800$ suggests that balancing can be naturally achieved. The observed typical FRB frequencies fall in the range of two orders of magnitude from $\sim 100$ MHz to $\sim 10$ GHz. This allows the typical Lorentz factor to vary within one order of magnitude among different bursts, so the value is not fine-tuned.

In order to compare the total observed isotropic luminosity via ICS and CR, we fix the emission radius to $500R_\star$ and normalize the net charge factor to be unity.
The most conservative estimate gives the transverse size of one bunch as $\gamma\lambda_{\rm frb}$
The total number of the charged bunches in the typical emission region can be estimated as 
\begin{equation}\label{eq:total number of bunch}
N_b\simeq\frac{\rho/\gamma}{\lambda_{\rm frb}}\frac{r/\gamma}{\gamma\lambda_{\rm frb}}\simeq 5.4\times10^{5} \ f r_{8.7}^2\gamma_{2.9}^{-3}\nu_9^{2},
\end{equation}
where $\rho=f r$ is the curvature radius of the dipole magnetic field, $f$ is a factor (here we normalize the value of $f$ to unity) connecting $\rho$ and radius $r$ which can be calculated as
\begin{equation}
f=\frac{\rho}{r}\simeq \frac{4}{3\sin\theta_0} \  \ {\rm for} \ \  \theta_0\lesssim0.5,
\end{equation}
which can reach $\gtrsim10$ when $\theta_0=0.1$, $\theta_0$ is the angle between the radius and magnetic axis.
The incident magnetosnic wave amplitude is fixed to $E_{\rm fmw,0}=10^{9} \ \rm esu$ at the magnetar surface for the ICS model.
The total received luminosity via ICS through 
coherent superposition can be calculated as
\begin{equation}
\begin{aligned}
L_{\rm ICS,coh}&\simeq\gamma^4 N_{b}^2N_{e,b}^2P_{\rm ICS}\\
&\simeq (3.5\times10^{42} \ {\rm erg \ s^{-1}}) \ f^2\gamma_{2.9}^6\omega_{i,4}^2r_{8.7}^{2}\nu_9^{-2}\zeta^2 \\
&\times P^{-2}R_{\star,6}^{2} E_{\rm fmw,0,9}^2,
\end{aligned}
\end{equation}
which is consistent with the FRB observations. Here the $\gamma^4$ factor comes from two parts \citep{Kumar2017}: 1. The observed luminosity of an individual emitting lepton is boosted by a factor of $\gamma^2$ since the observer time is modified by $1-\beta\cos\theta_v\sim 1/\gamma^2$ with $\theta_v<1/\gamma$. 2. This radiation is beamed within a $1/\gamma$ cone, thus the observed isotropic luminosity is a factor of $\gamma^2$ larger. 
Notice again that we have assumed that emission from different bunches are superposed coherently, which is a natural consequence for low-frequency-wave-induced bunches.

\subsection{Alf\'ven waves cannot be upscattered via inverse Compton scattering in the open field line region}

{A related question is whether kHz Alf\'ven waves can be also upscattered to GHz waves. Here we point out that this is impossible in the open field line regions. This is because the Alf\'ven waves and the leptons are moving in the same direction along magnetic field lines so that there is a Doppler deboost rather than Doppler boost. More generally, the incident waves frequency can be written in the comoving frame of a relativistic plasma as
\begin{equation}
\begin{aligned}
\omega'&=\omega_i\gamma(1-\beta\cos\theta_i)\\
&\simeq\left\{
\begin{array}{ll}
\omega_i/(2\gamma), &{\theta_i=0, \ \rm for \ {Alfven} \ waves},\\
\gamma\omega_i, &{\theta_i\gg0, \ \rm for \ fast \ magnetosonic \ waves}.
\end{array}
\right.
\end{aligned}
\end{equation}
In the lab frame, the scattered wave frequency is boosted by another factor of $\sim \gamma$. While the frequency of the scattered fast magnetosonic waves is boosted by a factor of $\sim \gamma^2$ from the low-frequency waves, that of the scattered Alf\'ven waves remains the same as the incident one. As a result, we do not consider the ICS effect of Alfv\'en waves in this paper. 
}

\section{Confronting coherent ICS model with observations}\label{sec:confront}

In this section, we will investigate the basic expected features of the coherent ICS model and confront these features with FRB observations.

\subsection{Polarization properties}\label{sec:polarization}

In this subsection, we outline a model for the ICS polarization properties and discuss the orthogonal modes (X-mode and O-mode) of the radiation. 
The polarization properties of a quasi-monochromatic electromagnetic wave can be described by four Stokes parameters \citep{Rybicki&Lightman1979}
\begin{equation}\label{Stokes}
\begin{aligned}
&I=\frac{1}{2}(E_x^{*}E_x+E_y^{*}E_y), \ &&Q=\frac{1}{2}(E_x^{*}E_x-E_y^{*}E_y),\\
&U={\rm Re}(E_x^{*}E_y), \ &&V={\rm Im}(E_x^{*}E_y),
\end{aligned}
\end{equation}
where $E_x$ and $E_y$ are the electric vector amplitudes of two linearly polarized wave eigen-modes perpendicular to the line of sight, the superscript $``*"$ denotes the conjugation of $E_{x/y}$, $I=\vert \Vec{E} \vert^2$ defines the total intensity, $Q$ and $U$ define linear polarization and its polarization angle ${\rm PA}=(1/2)\arctan(U/Q)$, and $V$ describes  circular polarization. The linear, circular, and the overall degree of polarization are described by $\Pi_{L}=(Q^2+U^2)^{1/2}/I$, $\Pi_{V}=V/I$ and $\Pi_{P}=(Q^2+U^2+V^2)^{1/2}/I$, respectively.
We only consider the electric field of the scattered waves due to $\vec E \times \vec B$ drift motion ($y$-component in Eq.(\ref{eq:bunch formation velocity})). 
We define the unit vector of line of sight as $\hat{n}'=\sin\theta_v'\cos\phi_v'\hat{x}+\sin\theta_v'\sin\phi_v'\hat{y}+\cos\theta_v'\hat{z}$.
Then the electric field of the scattered waves for one charged particle in the comoving frame can be calculated as
\begin{equation}\label{eq:single particle's field}
\begin{aligned}
\vec E_{\rm rad}'&\simeq\frac{q^2E_{\rm fmw}'\sin(\omega_i't')}{m_ecR'}\left(\frac{\omega_i'}{\omega_B'}\right)[\cos\phi_v'\sin\phi_v'\sin^2\theta_v'\hat{x}\\
&-(\cos^2\theta_v'+\cos^2\phi_v'\sin^2\theta_v')\hat{y}+\cos\theta_v'\sin\phi_v'\sin\theta_v'\hat{z}].
\end{aligned}
\end{equation}
We note that the value of the electric field of the scattered waves is determined by the projection of $\dot{\vec \beta}'$ onto the plane which is perpendicular to the line of sight, and there only exists one single oscillation frequency for the radiation field.
Therefore, when the incident wave is 100\% linearly polarized, the ICS radiation is also 100\% linear polarized in any viewing angle for one single particle. Non-zero circular polarization can be only generated via coherent superposition of many charges in a bunch by considering an asymmetry introduced by the incident waves and the curved field lines \citep{Qu&Zhang2023}.

For the on-axis case, i.e. $\theta_{v}=0$, the ICS radiation is always 100\% linearly polarized ($\Pi_L=100\%$) and no circularly polarized waves ($\Pi_V=0$) are produced \citep{Qu&Zhang2023}. 
In order to produce non-zero circular polarization, one needs to consider the asymmetry introduced by incident waves on a charged bunch. 
The incident waves will have a phase delay when encountering different particles distributed in the extended bunch.  
Therefore, circular polarization can be generated from the superposition of the scattered waves with different phases and polarization angles. 
When $\phi_v=0$, the line of sight lies in the $x-z$ plane which is the symmetric axis of the charged bunch, thus the radiation is $100\%$ linearly polarized waves. 
When $\phi_v\neq0$, the circular polarization will show up at an off-axis viewing angle, i.e. $\theta_v\neq0$. A larger $\theta_v$ will give a larger circular polarization degree at a specific $\phi_v$. 
The maximum circular polarization degree could reach $\sim40\%$, $\sim90\%$ and $\sim100\%$ for $\phi_v=10^\circ, 30^\circ$ and $60^\circ$, respectively, which are still within the main radiation cone ($\theta_v<1/\gamma$) \citep{Qu&Zhang2023}. 
When considering a broad radiation region inside the magnetosphere including more than one bunch, the superposition of the scattered waves leads to a high linear polarization in most cases since the radiation of most bunches is likely confined within the $1/\gamma$ cone and the circular polarized components cancel out each other. This might explain the fact that most FRBs from active repeaters possess high degrees of linear polarization ($\sim100\%$, \citealt{Jiang22}). Only under special geometries can a significant circular polarization show up. Observationally, the fraction of highly-circularly-polarized bursts are rare \citep{Jiang22}. On the other hand, very high degree of circular polarization 
as observed (J.-C. Jiang et al. 2024, in preparation) can be reproduced within the framework of the ICS model under rare geometric configurations.

In the following, we will provide the argument that both X-mode and O-mode can be generated through the ICS model.
Based on the electric field of a single particle's scattered waves  (Eq.(\ref{eq:single particle's field})), we can calculate the projection of the radiation electric field onto the $\vec k'-\vec B_0$ plane as
\begin{equation}\label{eq:criterion}
\begin{aligned}
\left|\vec E_{\rm rad}'\cdot(\hat{n}'\times\vec B_0)\right|&\sim \frac{qE_{\rm fmw}'}{m_ec}\left(\frac{\omega_i'}{\omega_B'}\right)B_0\cos\phi_v'\sin\theta_v'\\
&\left\{
\begin{aligned}
&=0 \ \& \ \theta_v'=0  \Rightarrow {\rm Mode \ coupling}, \\
&=0 \ \& \ \theta_v'\neq0 \Rightarrow {\rm O-mode}, \\
&\neq 0 \ \& \ \theta_v'\neq0 \Rightarrow {\rm X/O-mode},
\end{aligned}
\right.
\end{aligned}
\end{equation}
and the projection of electric field component along $\vec B_0$ as
\begin{equation}\label{eq:pureXmode}
\begin{aligned}
\vec E_{\rm rad}'\cdot\vec B_0&\sim \frac{qE_{\rm fmw}'}{m_ec}\left(\frac{\omega_i'}{\omega_B'}\right)B_0\cos\theta_v'\sin\phi_v'\sin\theta_v'\\
&\left\{
\begin{aligned}
&=0 \ \& \ \theta_v'=0 \ \Rightarrow \ {\rm Mode \ coupling}, \\
&=0 \ \& \ \theta_v'\neq0 \ \Rightarrow \ {\rm X-mode}, \\
&\neq0 \ \& \ \theta_v'\neq0 \ \Rightarrow \ {\rm X/O-mode},
\end{aligned}
\right.
\end{aligned}
\end{equation}
which can be used to determine the states of the radiation orthogonal modes, where the scattered wave vector has the same direction as the line of sight $\hat{n}'$. 
For the case of $\theta_v'=0$, the wave vector is along $\vec B_0$. The O-mode is coupled with the X-mode and it is not possible to distinguish between X- and O-modes. 
For the case of $\theta_v'\neq0$, the value of Eq.(\ref{eq:criterion}) equals to zero when $\phi_v'=\phi_v=0$ and $\pi$, and the radiation is O-mode. Notice that the line of sight lying in the $\vec E_{\rm fmw}'-\vec B_0$ plane can only give O-mode. Otherwise the radiation includes both X-mode and O-mode. When $\phi_v'=\pi/2$ and $3\pi/2$, the value of Eq.(\ref{eq:pureXmode}) equals to zero and only X-mode can be generated in the $\hat{n}'-\vec B_0$ plane.
Therefore, the observer can detect the superposed radiation of both X- and O-modes.

\subsection{Radiation spectra}\label{sec:spectra}

Motivated by recent observations of narrow spectra at least for some bursts \citep{ZhouDJ2022,ZhangYK2023,Sheikh2024}, we discuss the general expected features of the ICS spectra.

\begin{figure*}
\begin{center}
\begin{tabular}{ll}
\resizebox{94mm}{!}{\includegraphics[]{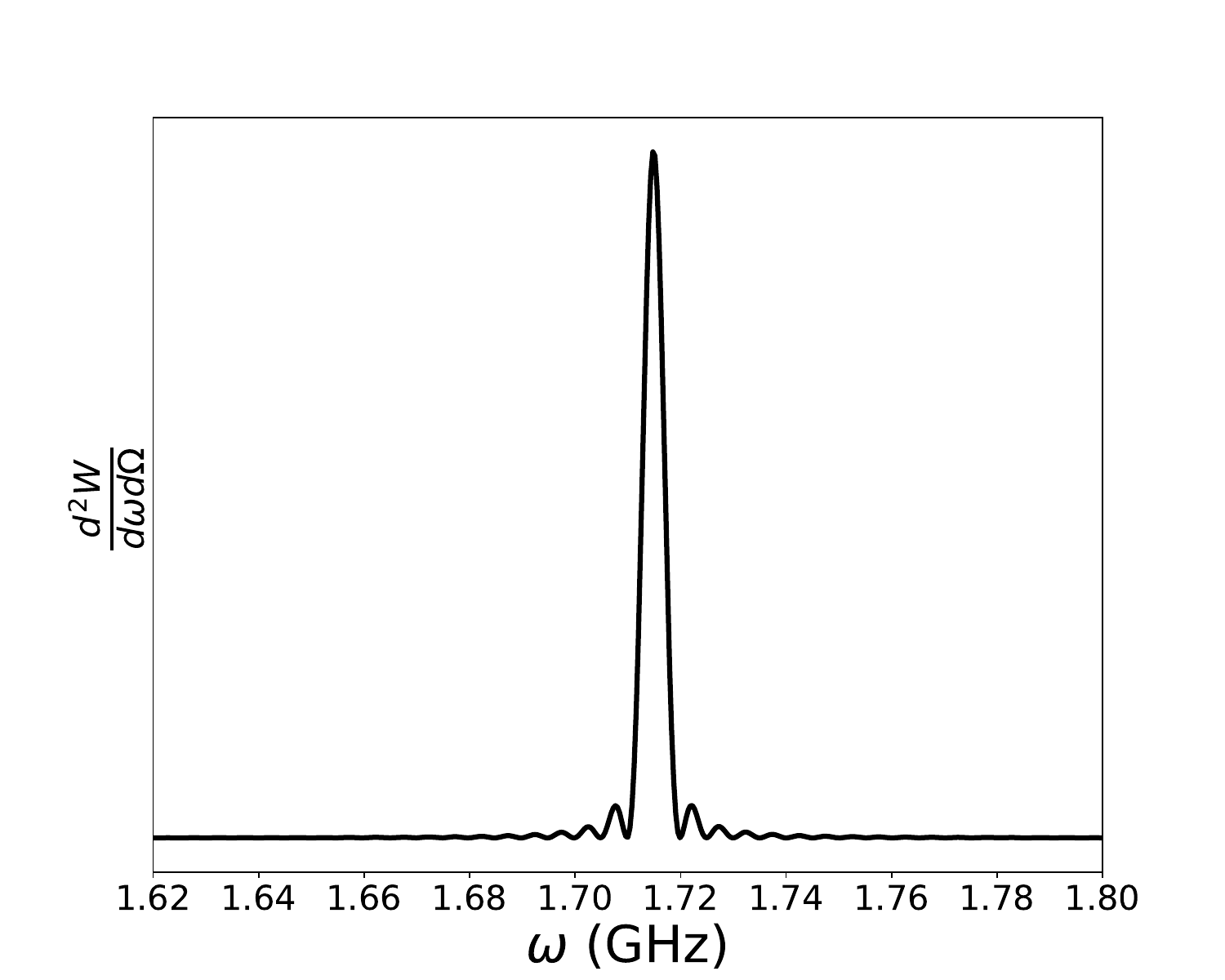}}&
\resizebox{91mm}{!}{\includegraphics[]{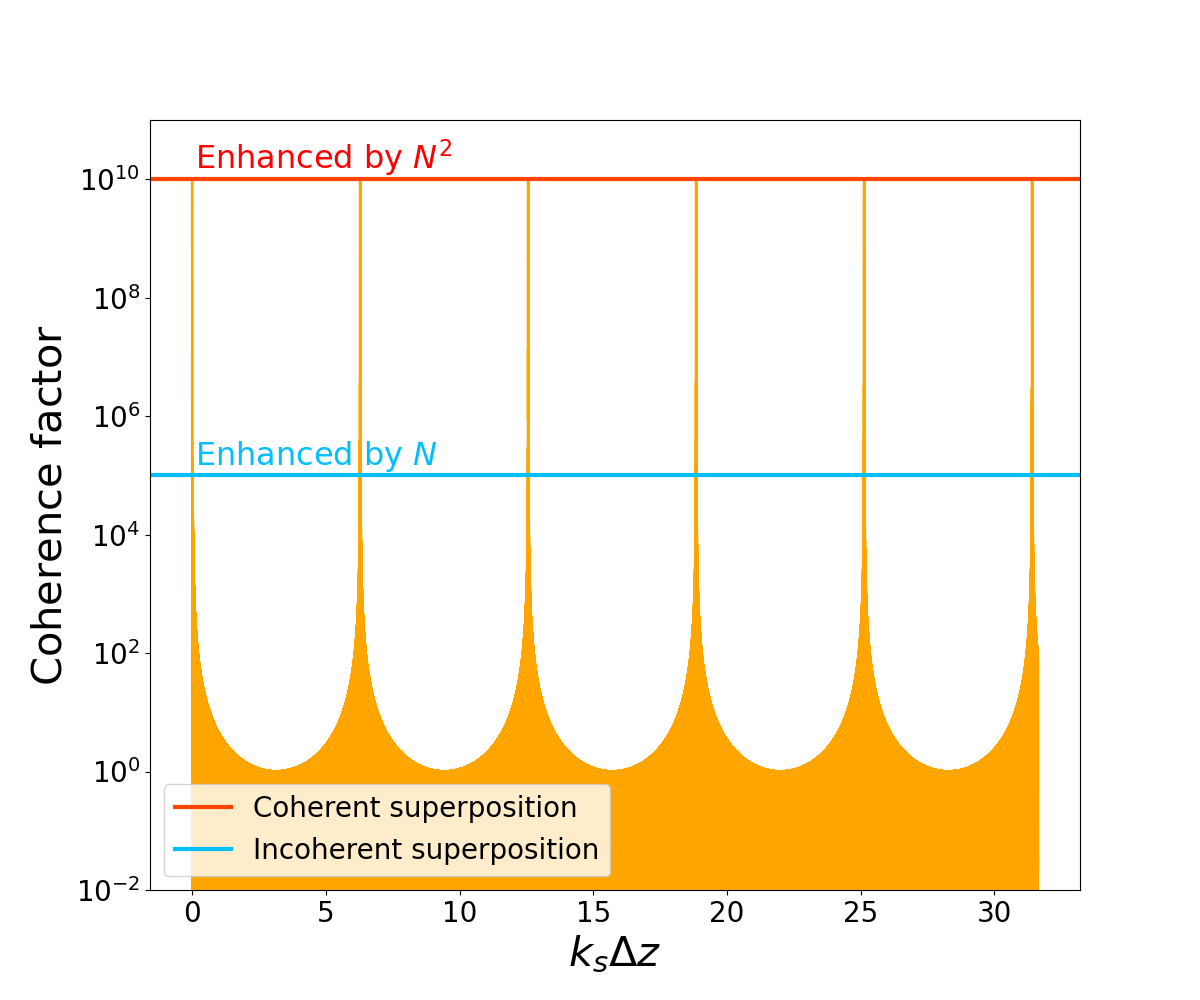}}
\end{tabular}
\caption{Left panel: The single bunch radiation spectra (solid black line) via ICS as a function of angular frequency $\omega$. Following parameters are adopted: Incident wave angle $\theta_i=30^\circ$, incident wave strength parameter $a=3.5\times10^9$, the incident wave frequency $\omega_i=10^4$ Hz, the bunch Lorentz factor $\gamma=800$ and the viewing angle $\theta_v=0$. Right panel: The coherence factor as a function of $k_s\Delta z$.}
\label{fig:coherent narrow}
\end{center}
\end{figure*}

(i) The ICS spectrum of a single charged particle with a fixed Lorentz factor scattering off a monochromatic low-frequency wave would give rise to a narrow spectrum. 
For a single charged particle, the energy radiated per unit solid angle per unit frequency interval is given by \citep{Rybicki&Lightman1979,Jackson1998}
\begin{equation}\label{eq:spectrum of single particle}
\begin{aligned}
\frac{d^2W}{d\omega d\Omega}=\frac{e^2\omega^2}{4\pi^2c}\left|\int_{-\infty}^{+\infty}{\hat{n}\times(\hat{n}\times{\vec\beta})}{\rm exp}([i\omega(t-\hat{n}\cdot\vec r/c)]dt \right|^2,
\end{aligned}
\end{equation}
where $\hat{n}$ is the unit vector along the wave propagation direction, i.e. along the line of sight. Assuming a monochromatic spectrum for the incident waves ($\omega_i=10^4 \ {\rm rad \ s^{-1}}$), we present the radiation spectra in the left panel of Fig. \ref{fig:coherent narrow}. One can see that the spectrum of the upscattered wave is roughly a $\delta$-function. This is because in the linear regime, the scattering is essentially Thomson scattering, i.e. in the comoving frame the particle is oscillating non-relativistically in the field of the incoming low frequency waves. The upscattered spectrum is peaked at $\nu_i' \sim\gamma \nu_i$. The spectrum would broaden in the non-linear regime when the particles move with a relativistic speed in the comoving frame. However, this regime is not achieved in the inner magnetosphere due to the suppression of cross section by a factor of $(\omega'/\omega_B')^2$ \citep{QKZ}. For the typical parameters adopted in the above discussion, one is always in the linear regime.

(ii) The spectrum can be broadened by two factors. First, the current assumption is that the incident low frequency waves have a narrow spectrum around $\nu_i$. 
It is envisaged that the low frequency waves are induced by seismic oscillations due to crustal deformations during the magnetar flare events. The exact spectra of such oscillations are unknown, and some work describes the spectra as power law \citep{Kasahara1981} shape, with the power law index very uncertain. This would affect the ICS spectrum. The monochromatic approximation is reasonably good if the power law index is very steep. 
In principle, the ICS model can also explain relatively broad-band spectra of apparently non-repeating FRBs, if the incident low-frequency waves bandwidth is broader or the emitting particles have a broader Lorentz factor distribution. However, other models can also account for these observations, so we keep our mind open for other mechanisms and even other sources to interpret apparently non-repeating FRBs.

(iii) Second, the spectrum can be broadened if the bunch Lorentz factor has a wide distribution. However, as discussed in \S\ref{sec:Eparallel}, the existence of $E_\parallel$ in the emission region would make leptons emit in the radiation reaction limited regime, so that a narrow particle energy spectrum is expected.

(iv) Even if the spectrum from an individual bunch is broader than that calculated in the left panel of Fig. \ref{fig:coherent narrow}, the observed spectrum can be further narrowed through a coherent superposition of emission from many bunches. As discussed in \S\ref{sec:bunching}, the incident waves may re-distribute the plasma along the magnetic field so that the emission from different bunches could be superposed coherently. 
The total radiation spectra should be modified by multiplying Eq.(\ref{eq:coherent factor}), see also \citep{Yang2023,Wang2023}. The coherence factor as a function of the phase is presented in the right panel of Fig. \ref{fig:coherent narrow}. 

(v) Finally, the ICS model has a specific prediction that could be tested with future observations. Besides the main emission component with a narrow spectrum, weaker but narrow emission components at higher harmonic components are expected. This is expected at both the single particle emission level (harmonic motion of particles in a finite time series) and the collective emission level (many bunches emit coherently with a proper spatial separation). Such a prediction could be tested with bright bursts detected with wide-band sensitive telescopes.

Observations show that active repeaters have narrow spectra with typical values of $\Delta \nu /\nu_0 \sim (0.2-0.3)$ \citep{ZhouDJ2022,ZhangYK2023}. Such a feature defies most of the models, but could be accommodated within the framework of the ICS model.

\section{Comparison between coherent CR and ICS models}\label{sec:comparison}
In this section, we compare the features of the ICS model with the widely discussed CR model, including common features and key differences.
\begin{itemize}
\item Both mechanisms apply inside the magnetar magnetosphere, thus sharing some predictions that are consistent with observational data: (i) The PA swing is expected within the rotation vector model as the line of sight sweeps across different magnetic field lines as the magnetar rotates. The PA evolution can also be flat if the magnetar spins slowly and if the emission region is far from the surface.  PA jumps might happen due to coherent or incoherent superposition of radiation fields.
(ii) The nanosecond variability timescale poses a stringent constraint on the emission radius, i.e.
$\delta t_{\rm var}\simeq r/(2c\Gamma^2)$. This favors magnetospheric models with emission radius within light cylinder and a high Lorentz factor of the emitting plasma. (iii) The magnetosphere models can overcome the generic constraints posed from the high-latitude emission effect of a spherically emitting shell \citep{KQZ2024}.

\item Emission frequency: The characteristic emission frequency of CR is $\propto \gamma^3/\rho$, which depends on $\gamma^3$ and the curvature radius $\rho$, 
while that of ICS is $\propto \gamma^2\omega_i$, which depends on $\gamma^2$ and incident wave frequency $\omega_i$. In order to produce GHz waves within the magnetosphere, relativistic particles are required for both mechanisms with Lorentz factors a few hundreds.

\item Emission power: The emission power of the ICS mechanism becomes dominant against the CR mechanism when the radius is greater than $\sim 100 R_\star$, because the cross section of the ICS rapidly increases with radius due to less suppression as the background magnetic field drops quickly with radius.

\item Bunch formation: For the CR mechanism, the formation of bunches requires an additional process, e.g. two stream plasma instabilities, and there is no explanation why the formed bunches have the size of the GHz wave wavelength (i.e. 30 cm). The ICS mechanism can naturally bunch a plasma to the desired size by the incident low-frequency waves to power the GHz waves.

\item Polarization properties: 
For emission of a single lepton, curvature radiation can produce both linear and circular polarization depending on the viewing angle, whereas ICS can only produce linear polarization if the incident wave is 100\% linear polarized. For emission from a charged bunch, the polarization properties are the same as the single particle case for CR. High circular polarization is achievable for an off-axis observer. 
For ICS, the case for a charged bunch can be different from the case of a single particle because the direction of the incoming low-frequency waves and the curved magnetic field lines introduce asymmetry in the system. Circular polarization can be generated because of 
the different phases and different polarization angles of the incoming waves for different particles within the bunch.
The ICS mechanism can achieve a higher circular polarization degree than CR.

Both CR and ICS mechanisms originate from the magnetosphere. So they all allow PA sweeping with the potential application of the rotating vector model. However, because FRBs are much brighter than radio pulses and are often associated with significant particle outflow and high-energy emission, the simple dipolar geometry may be significantly altered. This explains more diverse PA evolution behaviors of FRBs than pulsars.

\item Spectral width: For one single charged particle, the CR spectrum is intrinsically broad with $\Delta\nu/\nu_0>1$. In the high frequency regime ($\omega>\omega_c$, $\omega_c=3\gamma^3c/(2\rho)$ is the characteristic frequency) the intensity drops exponentially as $\sim e^{-\omega/\omega_c}$, but in the low frequency regime ($\omega<\omega_c$), the spectrum intensity is proportional to $\omega^{2/3}$, causing broadband emission. The ICS spectrum for a single particle is much narrower (nearly a $\delta$-function) if the incident waves are monochromatic. The nearly mono-energetic distribution of particle energy within the bunch also maintains the narrow spectrum. Even if the incident waves have a broader band, the spatial coherence among bunches induced by the low-frequency waves can further narrow down the emission spectrum. It is expected that the ICS mechanism tends to generate narrower spectra than the CR mechanism.

\item Required net charge factor: The emission powers of a single particle for the two mechanisms are presented in Fig. \ref{fig:emission power contour}. One can see that the ICS emission power can become greater than that of CR at larger radii. For the specific amplitude of incident waves we choose, the transition happens at $r > 100 R_\star$. As a result, the required degree of coherence is lower for ICS than CR. For typical parameters, the CR process usually requires the net charge factor to be several hundred times of the Goldreich-Julian density. This requires a high electron-positron pair density, which tends to screen $E_\parallel$ required in the model. The ICS model, on the other hand, only requires a fluctuation in the density and can even be in the sub-Goldreich-Julian regime to satisfy the observed luminosity. The plasma can be a single-specie one and the emission region is charge depleted. Such a picture is self-consistent with the requirement of an $E_\parallel$, whose origin and the concrete amplitude are subject to further study.

\end{itemize}

\section{Conclusion}\label{sec:conclusion}

In this paper, we revisit the idea that FRB emission is produced within the magnetar magnetosphere through an inverse Compton scattering process by relativistic particles off low-frequency X-mode waves in the form of fast magnetosonic waves \citep{Zhang22}.  
We have investigated such a physical process in greater detail and found that it is an attractive mechanism that can account for broad FRB phenomenology. The main findings of our paper include the following:
\begin{itemize}
\item Crustal oscillations during a flaring event of the magnetar can excite kHz low-frequency waves in the form of Alfv\'en waves and possibly in fast magnetosonic waves as well. Fast magnetosonic waves can be also generated via conversion from Alfv\'en waves with the conversion efficiency increases with altitude. At the emission region (typically hundreds of neutron star radii), the amplitude of fast magnetosonic waves can be large enough to power ICS emission by relativistic particles. The kHz low-frequency waves are boosted to GHz through ICS to power FRBs.
\item By adopting the correct X-mode cross section, the ICS power for individual particles is much smaller than that estimated in \cite{Zhang22}, and increases rapidly with radius as the magnetic suppression of the cross section becomes less important. For reasonable parameters, the ICS power starts to exceed the CR power and becomes dominant to produce coherent emission at $r \gtrsim 100 R_\star$, and can be several orders of magnitude higher than the CR power, greatly reducing the required degree of coherence. The FRB luminosity is achievable even with a sub-Goldreich-Julian plasma density. 
\item The incident low-frequency fast magnetosonic waves will modulate the motion of the particles through the 
electric field force and the $\vec E_{\rm fmw}\times\vec B_0$ drift. Such modulations will redistribute an initially fluctuating plasma to naturally form spatial bunches with the right size to power the GHz FRB emission.
\item To sustain high emission power to meet the FRB luminosity, the bunches are required to be continuously accelerated by a $E_\parallel$, which is a natural consequence of charge depletion in the emission region.
\item This model can reproduce key FRB observational features, including high degrees of both linear and circular polarization, diverse behavior of polarization angle sweeping, as well as the very narrow spectra observed in many bursts of repeating FRBs. 
\end{itemize}

In view of the mounting evidence suggesting a magnetospheric origin of FRB emission at least for repeating sources and the advantages of the ICS model over the traditional CR model (e.g. high emission power, the ability to self-bunch due to the low-frequency waves, and the ability of producing narrow spectra and high circular polarization), we argue that the ICS mechanism is likely responsible to the FRB emission of repeating FRB sources within the framework of the magnetar central engine. 

\section*{Acknowledgements}
We thank Omer Blaes, Alexander Y. Chen, Kunihito Ioka, Pawan Kumar, Wenbin Lu, Lorenzo Sironi and Yuanpei Yang for helpful discussion and comments. This work is supported by Nevada Center for Astrophysics, NASA 80NSSC23M0104 and a Top Tier Doctoral Graduate Research Assistantship (TTDGRA) at University of Nevada, Las Vegas.

\appendix

\section{Dispersion relations of low frequency waves in a cold plasma and three-wave interactions}\label{App:dispersion relation}

In this Appendix, we present a brief derivation of the dispersion relation of electromagnetic wave modes in a cold plasma. In an environment of a magnetar magnetosphere, we consider the propagation of a weak electromagnetic wave with perturbation electric field $\vec E_1$ in a background non-relativistic plasma mainly consists of electron-positron pairs. In the linear regime, the particle's motion equation can be written in the Fourier space as
\begin{equation}
-i\omega m_s \vec v_{1s}=q_s\left(\vec E_1+\frac{1}{c}\vec v_{1s}\times \vec B_0\right),
\end{equation}
where the background magnetic field $\vec B_0$ is chosen to be aligned with $z$-axis and the subscript $s$ denotes sth species. We define $\omega_{Bs}=q_s B_0/(m_s c)$ as the cyclotron frequency. The $xyz$-components of the motion equation can be written as
\begin{equation}
x: \ -i\omega v_{xs}-\omega_{Bj}v_{ys}=\frac{q_s E_x}{m_s}.
\end{equation}
\begin{equation}
y: \ -i\omega v_{ys}+\omega_{Bs}v_{xs}=\frac{q_s E_y}{m_s}.
\end{equation}
\begin{equation}
z: \ -i\omega v_{zs}=\frac{q_s E_z}{m_s} \ \Rightarrow \ v_{zs}=\frac{iq_sE_z}{m_s\omega}.
\end{equation}
The perturbation velocity of the $xy$-components can be solved as
\begin{equation}\label{eq:v_{xs}}
v_{xs}=\frac{iq_s}{m_s(\omega^2-\omega_{Bs}^2)}(\omega E_x+i\omega_{Bs}E_y).
\end{equation}
\begin{equation}\label{eq:v_{ys}}
v_{ys}=\frac{iq_s}{m_s(\omega^2-\omega_{Bs}^2)}(-i\omega_{Bs} E_x+\omega E_y).
\end{equation}
We introduce the rotating coordinates, i.e. $v_{\pm}=v_{xs}\pm iv_{ys}$ and $E_{\pm}=E_x\pm iE_y$, then we have
\begin{equation}
v_{\pm}=\frac{iq_s E_{\pm}}{m_s(\omega\mp\omega_{Bs})}.
\end{equation}
The linear relation for current and susceptibility is
\begin{equation}
\vec j_s=\vec \sigma_s\cdot\vec E=-\frac{i\omega}{4\pi}\vec\chi_s\cdot\vec E=q_s n_s \vec v_s,
\end{equation}
where $\vec \chi_s$ is the susceptibility. 
For the $z$-component and the $xy$ components, we have
\begin{equation}
\chi_{zz,s}=-\frac{\omega_{ps}^2}{\omega^2}, ~~ \ \& ~~ \ \chi_s^{\pm}=-\frac{\omega_{ps}^2}{\omega(\omega\mp\omega_{Bs})},
\end{equation}
where $\omega_{ps}^2=4\pi q_s^2 n_s/m_s$ is the plasma frequency for species s.
Now we can apply the current expressions to solve for the cold plasma dielectric tensor:
\begin{equation}
D_{k}=\epsilon_{kl}E_l \ \& \ D_k=E_k+\frac{4\pi i}{\omega}j_k \ \Rightarrow \ \epsilon_{kl}E_l=E_k+\frac{4\pi i}{\omega}j_k.
\end{equation}
The dielectric tensor can be written as
\begin{equation}
\vec\epsilon\cdot\vec E=\left( 
  \begin{array}{ccc}  
    S & -iD & 0\\
    iD & S & 0\\
    0 & 0 & P\\
  \end{array}
\right)\left( 
  \begin{array}{ccc}  
    E_x\\
    E_y\\
    E_z\\
  \end{array}
\right),
\end{equation}
where the quantities $S,D$ and $P$ are defined as
\begin{equation}
S=\frac{1}{2}(R+L), \ D=\frac{1}{2}(R-L).
\end{equation}
\begin{equation}
R=1-\sum_s \frac{\omega_{ps}^2}{\omega(\omega+\omega_{Bs})}, \ L=1-\sum_s \frac{\omega_{ps}^2}{\omega(\omega-\omega_{Bs})}.
\end{equation}
\begin{equation}
P=1-\sum_s\frac{\omega_{ps}^2}{\omega^2}.
\end{equation}
Notice that the dielectric tensor is additive in its components.
The wave equation is given by
\begin{equation}
\nabla\times(\nabla\times\Vec{E})+\frac{1}{c^2}\frac{\partial^2\Vec{E}}{\partial t^2}=-\frac{4\pi}{c^2}\frac{\partial\Vec{j}}{\partial t}.
\end{equation}
We recall that the oscillating field quantities are assumed to vary as ${\rm exp}(i\Vec{k}\cdot\Vec{r}-i\omega t)$, i.e. the plane wave approximation ($\nabla\rightarrow i\Vec{k}$ and $\partial/\partial t\rightarrow-i\omega$), and consider the homogeneous part. We define the refractive index $\vec n={\Vec{k}c}/{\omega}$.
The wave equation can be re-written as
\begin{equation}
\Vec{n}\times(\Vec{n}\times\Vec{E})+\Vec{\epsilon}\cdot\Vec{E}=0.
\end{equation}
We define the angle $\theta$ between the background magnetic field $\Vec{B}_0$ (along $z$-axis) and the wave vector $\Vec{k}$ (in the $x-z$ plane). One can see that $\vec n$ also lies in the $x-z$ plane, i.e. $\vec n=(n\sin\theta,0,n\cos\theta)$. We can then expand $\vec n\times(\vec n\times\vec E)$ in terms of matrix as
\begin{equation}
\vec n\times(\vec n\times\vec E)=\left( 
  \begin{array}{ccc}  
    -n^2\cos^2\theta & 0 & n^2\cos\theta\sin\theta\\
    0 & -n^2 & 0\\
    n^2\cos\theta\sin\theta & 0 & -n^2\sin^2\theta\\
  \end{array}
\right)\left( 
  \begin{array}{ccc}  
    E_x\\
    E_y\\
    E_z\\
  \end{array}
\right).
\end{equation}
Then the wave equation can be re-written as
\begin{equation}\label{eq:wave equation}
\left( 
  \begin{array}{ccc}  
    S-n^2\cos^2\theta & -iD & n^2\cos\theta\sin\theta\\
    iD & S-n^2 & 0\\
    n^2\cos\theta\sin\theta & 0 & P-n^2\sin^2\theta\\
  \end{array}
\right)\left( 
  \begin{array}{ccc}  
    E_x\\
    E_y\\
    E_z\\
  \end{array}
\right)=0.
\end{equation}
The condition for a nontrivial solution of the wave equation is that the determinant of the $3\times3$ matrix is zero. In the first step, we start with the case that the wave frequency is much greater than the plasma frequency. Then the plasma-related dielectric term $P\rightarrow 1$ and the electric field along the background magnetic field $E_z$ can be determined as
\begin{equation}
E_z\simeq \frac{n^2\cos\theta\sin\theta}{n^2\sin^2\theta-1}E_x.
\end{equation}
For the parallel propagation case ($\theta=0$), the O-mode is exactly the same as the X-mode and the corresponding wave electric field has no $z$-component, i.e. $E_z=0$. For the quasi-parallel case ($\theta\rightarrow0$), we can obtain $|E_z|\simeq \theta n^2E_x$ which is non-zero but is close to zero for O-mode waves.

We consider the low frequency limit case that the wave frequency is much smaller than the plasma frequency over all species, i.e. $\omega\ll\sum_s\omega_{ps}$. One can estimate the plasma-related dielectric term as
\begin{equation}
P=1-\frac{\omega_{p,+}^2+\omega_{p,-}^2}{\omega^2}\rightarrow -\infty.
\end{equation}
Thus the value of $E_z$ is determined as $(P-n^2\sin^2\theta)E_z\sim (-\infty)E_z$ which leads to $E_z\rightarrow 0$ for any propagation angles, i.e. the $z$-component of wave electric field is highly suppressed by plasma screening due to rapid oscillations.
To quantify how the plasma term ($P$) and the $\vec k-\vec B_0$ angle ($\theta$) influence the ratio of $E_z/E_x$, notice that the dispersion relation of O-mode for the quasi-perpendicular case can be written as
\begin{equation}
n_{\rm O}^2\simeq\frac{\omega^2-\omega_p^2}{\omega^2-\omega_p^2\cos^2\theta}.
\end{equation}
Then the ratio of $E_z/E_x$ can be calculated as
\begin{equation}
\frac{E_z}{E_x}={\tan\theta}\left(\frac{\omega_p^2}{\omega^2}-1\right)^{-1}.
\end{equation}
It should be pointed out that the wave vector is quasi-parallel to the background magnetic field ($\theta\rightarrow0$) and O-mode is exactly the same as X-mode, then $E_x=E_z=0$ and $n_{\rm O}^2=n_{\rm X}^2=1$. When $\theta=\pi/2$, the ratio $E_z/E_x\rightarrow\infty$ since $\vec k$ is along $x$-axis and $E_x=0$. In both cases ($\omega\gg\omega_p$ and $\omega\ll\omega_p$), the wave electric field is always perpendicular to $\vec B_0$.

Two general solutions for low frequency waves can be solved from Eq.(\ref{eq:wave equation}), which can be described below:
\begin{itemize}
\item The first solution has the dispersion relation by considering $E_x\neq0$ and $E_y=0$, then we have $S=n^2\cos^2\theta$ which can be re-written as
\begin{equation}
\omega^2=k^2V_A^2\cos^2\theta=k^2c^2\cos^2\theta.
\end{equation}
This is called Alf\'ven waves (also shear Alf\'ven waves), where the Alf\'ven velocity $V_A$ is close to the speed of light inside the magnetar magnetosphere. The waves' electric field is parallel to the $\vec k-\vec B_0$ plane which can be described as
\begin{equation}
\vec E_{\rm aw}=\frac{\vec k-k\cos\theta\hat{z}}{k\sin\theta}E_{\rm aw}.
\end{equation}
The waves' magnetic field and charge density can be obtained by making use of Maxwell's equations as
\begin{equation}
\vec B_{\rm aw}=\frac{c}{\omega}\vec k\times\vec E_{\rm aw}=-\frac{c}{\omega}\frac{\vec k\times\hat{z}\cos\theta}{\sin\theta}E_{\rm aw},
\end{equation} 
and
\begin{equation}
\nabla\cdot\vec E_{\rm aw}=4\pi\rho \ \Rightarrow \ \rho=\frac{i}{4\pi}E_{\rm aw}k \sin\theta.
\end{equation}

\item  We consider $E_x=0$ and $E_y\neq0$, then we have $S=n^2$ and the second solution can be written as
\begin{equation}
\omega^2=k^2V_A^2=k^2c^2.
\end{equation}
This is called fast magnetosonic waves (also called compressional Alf\'ven waves). The waves' electric field is perpendicular to the $\vec k-\vec B_0$ plane, which can be expressed as
\begin{equation}
\vec E_{\rm fmw}=\hat{B}_0\times\hat{k} E_{\rm fmw}=\frac{\hat{z}\times\vec k}{k\sin\theta}E_{\rm fmw}.
\end{equation}
One can see that $\vec k\cdot(\hat{z}\times\vec k)=0$. Thus the charge density $\rho=0$.
The waves' magnetic field can be obtained by making use of Maxwell's equations as
\begin{equation}
\vec B_{\rm fmw}=\frac{c}{\omega}\vec k\times\vec E_{\rm fmw}=\frac{k\hat{z}-\cos\theta \vec{k}}{k\sin\theta}E_{\rm fmw}.
\end{equation}

\end{itemize}

For completeness, in the following we  briefly discuss the conversion from Alf\'ven waves to fast magnetosonic waves which has been investigated by various authors \citep{Melrose&McPhedran1991,Yuan2021,Chen2024}. We claim that the conversion of Alf\'ven modes to magnetosonic modes is already efficient enough to satisfy the energy requirement for FRBs via the ICS process.
To the first order, the force-free MHD equations can be written as
\begin{equation}
\rho\vec E+\frac{1}{c}\vec j\times\vec B+\frac{1}{c}\vec j^{(1)}\times\vec B_0=0,
\end{equation}
and the Maxwell's equations can be written as
\begin{equation}\label{eq:curl of E}
i\vec k\times E^{(1)}=\frac{i\omega}{c}\vec B^{(1)}-\frac{1}{c}\frac{\partial \vec B}{\partial t}.
\end{equation}
We perform Fourier transformation to the force-free condition to obtain
\begin{equation}
\sum_{k'}\left(\rho_{k'}\vec E_{k-k'}+\frac{1}{c}\vec j_{k'}\times\vec B_{k-k'}\right)+\frac{1}{c}\vec j_k^{(1)}\times\vec B_0=0.
\end{equation}
One can see that the wave number in the Fourier space remains $k$. 
The current density can be calculated from the Ampere’s equation and we take the curl of Eq.(\ref{eq:curl of E}) to obtain
\begin{equation}\label{eq:current}
\begin{aligned}
\frac{4\pi}{c}\vec j^{(1)}&=i\vec k\times\vec B^{(1)}+\frac{i\omega}{c}\vec E^{(1)}-\frac{1}{c}\frac{\partial \vec E}{\partial t}\\
&=\frac{i}{\omega c}\left[(\omega^2-k^2c^2)\vec E^{(1)}+c^2(\vec k\cdot\vec E^{(2)})\vec k\right]\\
&+\frac{1}{\omega}\frac{\partial}{\partial t}\vec k\times\vec B-\frac{1}{c}\frac{\partial\vec E}{\partial t}.
\end{aligned}
\end{equation}
We again take the curl of $\vec E$ and Eq.(\ref{eq:current}) can be re-written as
\begin{equation}\label{eq:re-write current}
\begin{aligned}
\frac{4\pi}{c}\vec j^{(1)}&=\frac{i}{\omega c}\left[(\omega^2-k^2c^2)\vec E^{(1)}+c^2(\vec k\cdot\vec E^{(1)})\vec k\right]\\
&+\frac{c}{\omega^2}\left[\left(\vec k\cdot\frac{\partial\vec E}{\partial t}\right)\vec k-\left(\frac{\omega^2}{c^2}+k^2\right)\frac{\partial\vec E}{\partial t}\right].
\end{aligned}
\end{equation}
The energy conservation of three-wave interaction requires $\vec k_1+\vec k_2=\vec k$ and the force-free condition can be re-written as 
\begin{equation}
\rho_{k_2}\vec E_{k_1}+c^{-1}\vec j_{k_2}\times\vec B_{k_1}+c^{-1}\vec j_k^{(1)}\times\vec B_0=0
\end{equation}
for a single wave number $k_1$ initially.
Here we consider an Alf\'ven wave with $\vec k_1$ interacting with a time-independent background with $\vec k_2$ and generate a fast magnetosonic wave with $\vec k$. We substitute Eq.(\ref{eq:re-write current}) into the Fourier transformed force-free condition and obtain
\begin{equation}\label{eq:forcefree}
\begin{aligned}
&\rho_{k_{2}}\frac{\vec k_{\rm aw}-k_{\rm aw}\cos\theta_{\rm aw}\hat{z}}{k_{\rm aw}\sin\theta_{\rm aw}}E_{\rm aw}-\frac{1}{c}\vec j_{k_2}\times\frac{\vec k_{\rm aw}\times\hat{z}\cos\theta_{\rm aw}}{\sin\theta_{\rm aw}}E_{\rm aw}\\
&+\frac{1}{4\pi c}\left(-2\frac{\hat{z}\times\vec k_{\rm fmw}}{k_{\rm fmw}\sin\theta_{\rm fmw}}\frac{\partial E_{\rm fmw}}{\partial t}\right)\times\vec B_0=0,
\end{aligned}
\end{equation}
where we have applied $\vec k_{\rm fmw}\cdot \vec E_{\rm fmw}=0$ and the dispersion relation $\omega^2=k^2c^2$ for fast magnetosonic waves. Then we take a dot product with $\vec k_{\rm fmw}$ of Eq.(\ref{eq:forcefree}) to obtain \citep{Yuan2021}
\begin{equation}
\begin{aligned}
\frac{\partial E_{\rm fmw}}{\partial t}&=\frac{2\pi E_{\rm aw}\rho_{k_2} c}{B_0k_{\rm fmw}\sin\theta_{\rm fmw}}\frac{\vec k_{\rm aw}\cdot\vec k_{\rm fmw}-k_{\rm aw}\cos\theta_{\rm aw}\hat{z}\cdot\vec k_{\rm fmw}}{k_{\rm aw}\sin\theta_{\rm aw}}\\
&=\frac{E_{\rm aw}\Omega}{\sin\theta_{\rm fmw}}\frac{\cos(\theta_{\rm aw}-\cos\theta_{\rm fmw})-\cos\theta_{\rm aw}\cos\theta_{\rm fmw}}{\sin\theta_{\rm aw}}\\
&=E_{\rm aw}\Omega \ \Rightarrow \ E_{\rm fmw}\simeq E_{\rm aw}\frac{r}{R_{\rm LC}},
\end{aligned}
\end{equation}
where the charge number density $\rho_{k_2}$ follows the GJ-density inside the magnetosphere, $\Omega$ is the magnetar angular velocity, and $t\simeq r/c$ is the typical light travel time.
One can see that the fast magnetosonic wave amplitude is only related to the Alf\'ven wave amplitude and the magnetar rotation period.

\section{Electromagnetic field transformation for fast magnetosonic waves}\label{App:transformation}
In the lab frame, we consider the comoving frame of one charged particle with velocity $\vec v$ along $z$-axis.

We consider a general incident angle case that the propagation direction is in the $y-z$ plane with incident angle $\theta_i$ with respect to $z$-axis. The electric and magnetic fields can be written as
\begin{equation}
\vec E_w=E_{w,0}\sin(kr-\omega t)\hat{x},
\end{equation}
and 
\begin{equation}
\begin{aligned}
&\vec B_{w,\parallel}=-E_{w,0}\sin\theta_i\sin(kr-\omega t)\hat{z},\\
&\vec B_{w,\perp}=E_{w,0}\cos\theta_i\sin(kr-\omega t)\hat{y}.
\end{aligned}
\end{equation}
In the comoving frame, for the parallel components, we have $\vec E_\parallel'=\vec E_\parallel=0$ and
\begin{equation}
\vec B_\parallel'=\vec B_{w,\parallel}=-E_{w,0}\sin\theta_i\sin(kr-\omega t)\hat{z}.
\end{equation}
For the perpendicular components, we have
\begin{equation}\label{eq:transformed E-field}
\begin{aligned}
E_w'\hat{x}&=\gamma[E_{w,0}\sin(kz-\omega t)\hat{x}+\vec\beta\times\vec B_{w,\perp}]\\
&=\gamma(1-\beta\cos\theta_i)E_{w,0}\sin(kr-\omega t)\hat{x}\\
&=\frac{1}{\cal D}E_{w,0}\sin(kr-\omega t)\hat{x},
\end{aligned}
\end{equation}
where $\beta=\sqrt{1-\gamma^{-2}}$ is the dimensionless velocity of the charged particle and ${\cal D}=1/[\gamma(1-\beta\cos\theta_i)]$ is the Doppler factor, and
\begin{equation}
\begin{aligned}
B_w'\hat{y}&=\gamma[E_{w,0}\cos\theta_i \sin(kz-\omega t)\hat{y}-\vec\beta\times\vec E_w]\\
&=\gamma (\cos\theta_i-\beta) E_{w,0}\sin(kr-\omega t)\hat{y}.
\end{aligned}
\end{equation}

A special case is that the incident waves and background magnetic field are along $z$-axis. For the perpendicular components (perpendicular to $z$-axis), one has
\begin{equation}\label{eq:transformed E-field parallel}
\begin{aligned}
E_w'\hat{x}
&\simeq \frac{1}{2\gamma}E_{w,0}\sin(kz-\omega t)\hat{x}.
\end{aligned}
\end{equation}
One can see that the amplitude of the incident waves is de-boosted by a factor of $\sim2\gamma$ in the comoving frame of the relativistic plasma since ${\cal D}\simeq2\gamma$ for $\theta_i=0$ and $\beta\sim1$.

\section{Cross sections of X-mode and O-mode in highly magnetized plasma}\label{App:cross section}

In this appendix, we present a classical electromagnetic derivation on the cross sections of X-mode and O-mode waves in a highly magnetized plasma in a magnetar magnetosphere, which were investigated via a quantum treatment by \cite{Herold1979}. All calculations are done in the comoving frame of the relativistic plasma (all quantities in the comoving frame are denoted with a prime ($'$)).

The radiation electric and magnetic fields of one charged particle (here we consider one positron) in the comoving frame can be written as \citep{Jackson1998,Rybicki&Lightman1979}
\begin{equation}\label{eq:radiation E-field}
\vec E_{\rm rad}'=\frac{e}{cR'}[\hat n'\times(\hat{n}'\times\dot{\vec\beta}')] \ \& \ \vec B_{\rm rad}'=\hat{n}'\times\vec E_{\rm rad}'.
\end{equation}
We consider that the background magnetic field is $B_0\hat{z}$ and the perturbed wave electric field is $\vec E_x'=E_0'{\rm exp}(-i\omega' t')$ along $x-$axis. The effect of magnetic field component on the particle's motion is trivial in the field of $\vec B_0$. 
The velocity of the charged particle in the X-mode electromagnetic wave and the strong background magnetic field can be written as
\begin{equation}\label{eq:velocity}
\vec v'=\frac{ie\omega' E_x'}{m_e(\omega'^2-\omega_B'^2)}\hat{x}+\frac{e\omega_B' E_x'}{m_e(\omega'^2-\omega_B'^2)}\hat{y}.
\end{equation}
The cross section in the comoving frame of plasma can be calculated as
\begin{equation}
\sigma'=\frac{P'}{S'}=\frac{E_{\rm rad}'^2R'^2}{E_0'^2},
\end{equation}
where $P'$ and $S'$ are the emitted power and the incident Poynting flux.
We substitute Eq.(\ref{eq:velocity}) into Eq.(\ref{eq:radiation E-field}), then the energy per unit time radiated by one charged particle in the X-mode waves can be calculated as
\begin{equation}
\begin{aligned}
\sigma'({\rm X})&=\frac{8\pi}{3}\left(\frac{e^2}{m_ec^2}\right)^2\left[\frac{\omega'^4}{(\omega'^2-\omega_B'^2)^2}+\frac{\omega'^2\omega_B'^2}{(\omega'^2-\omega_B'^2)^2}\right] \\
&=\frac{\sigma_{\rm T}}{2}\left[\frac{\omega'^2}{(\omega'+\omega'_B)^2}+\frac{\omega'^2}{(\omega'-\omega'_B)^2}\right].
\end{aligned}
\end{equation}
In a highly magnetized environment, $\omega_B'\gg\omega'$ and the cross section of X-mode is $\sim\sigma_{\rm T}(\omega'/\omega_B')^2$. There exist two oscillation directions for the particle ($x$ and $y$, see Eq.(\ref{eq:velocity})) and both of them should contribute to the scattering radiation. However, in a highly magnetized environment, one can see that the $x$-direction component only contributes $(\omega'/\omega_B')^4$ to the cross section, which can be neglected, whereas the $y$-direction component contributes $(\omega'/\omega_B')^2$, which is the dominant term.

In the next step, we consider that the incident wave electric field has a non-zero component ($E_z'$) along $z$-axis, i.e. O-mode waves. The angle between the wave vector and $\vec B_0'$ is assumed to be $\theta_i'$. Then we can obtain the wave amplitude $|E_z'|=E_0'\sin\theta_i'$ and the velocity value along $z$-axis as
\begin{equation}\label{eq:v_z}
|v_z'|=\frac{ie|E_z'|}{m_e\omega'}=\frac{ieE_0'}{m_e\omega'}\sin\theta_i'.
\end{equation}
We substitute Eq.(\ref{eq:v_z}) into Eq.(\ref{eq:radiation E-field}) and the cross section due to the electric field along $z-$axis can be calculated as $\sigma'=\sigma_{\rm T}\sin^2\theta_i'$. The electric field perpendicular to $\vec B_0$ component is $|E_x'|=E_0'\cos\theta_i'$ and the cross section due to the $x$-component should be modified by multiplying $\cos^2\theta_i'$.
Thus the cross section of O-mode waves can be written as
\begin{equation}
\begin{aligned}
\sigma'({\rm O})&=\sigma_{\rm T}\sin^2\theta_i'+\sigma'({\rm X})\cos^2\theta_i'\\
&=\sigma_{\rm T}\left\{\sin^2\theta_i'+\frac{1}{2}\cos^2\theta_i'\left[\frac{\omega'^2}{(\omega'+\omega'_B)^2}+\frac{\omega'^2}{(\omega'-\omega'_B)^2}\right]\right\}.
\end{aligned}
\end{equation}
Noticing the first term $\sin^2\theta_i'$, and noticing that the plasma can only move along the strong background magnetic field, one can see that the radiation is not suppressed much for O-mode waves, i.e. $\sigma'({\rm O})\gg\sigma'({\rm X})$ due to the existence of $\sin^2\theta_i'$.
It should be pointed out that one cannot distinguish the X-mode and O-mode for the quasi-parallel propagation case ($\theta_i\rightarrow0$), i.e. $\sigma'({\rm X})=\sigma'({\rm O})$.


\begin{thebibliography}{}
\expandafter\ifx\csname natexlab\endcsname\relax\def\natexlab#1{#1}\fi
\providecommand{\url}[1]{\href{#1}{#1}}
\providecommand{\dodoi}[1]{doi:~\href{http://doi.org/#1}{\nolinkurl{#1}}}
\providecommand{\doeprint}[1]{\href{http://ascl.net/#1}{\nolinkurl{http://ascl.net/#1}}}
\providecommand{\doarXiv}[1]{\href{https://arxiv.org/abs/#1}{\nolinkurl{https://arxiv.org/abs/#1}}}

\bibitem[{{Bannister} {et~al.}(2019){Bannister}, {Deller}, {Phillips}, {Macquart}, {Prochaska}, {Tejos}, {Ryder}, {Sadler}, {Shannon}, {Simha}, {Day}, {McQuinn}, {North-Hickey}, {Bhandari}, {Arcus}, {Bennert}, {Burchett}, {Bouwhuis}, {Dodson}, {Ekers}, {Farah}, {Flynn}, {James}, {Kerr}, {Lenc}, {Mahony}, {O'Meara}, {Os{\l}owski}, {Qiu}, {Treu}, {U}, {Bateman}, {Bock}, {Bolton}, {Brown}, {Bunton}, {Chippendale}, {Cooray}, {Cornwell}, {Gupta}, {Hayman}, {Kesteven}, {Koribalski}, {MacLeod}, {McClure-Griffiths}, {Neuhold}, {Norris}, {Pilawa}, {Qiao}, {Reynolds}, {Roxby}, {Shimwell}, {Voronkov}, \& {Wilson}}]{Bannister2019}
{Bannister}, K.~W., {Deller}, A.~T., {Phillips}, C., {et~al.} 2019, Science, 365, 565, \dodoi{10.1126/science.aaw5903}

\bibitem[{{Beloborodov}(2017)}]{Beloborodov2017}
{Beloborodov}, A.~M. 2017, \apjl, 843, L26, \dodoi{10.3847/2041-8213/aa78f3}

\bibitem[{{Beloborodov}(2020)}]{Beloborodov2020}
---. 2020, \apj, 896, 142, \dodoi{10.3847/1538-4357/ab83eb}

\bibitem[{{Beloborodov}(2021)}]{Beloborodov2021}
---. 2021, \apjl, 922, L7, \dodoi{10.3847/2041-8213/ac2fa0}

\bibitem[{{Beloborodov}(2022)}]{Beloborodov2022}
---. 2022, \prl, 128, 255003, \dodoi{10.1103/PhysRevLett.128.255003}

\bibitem[{{Beniamini} \& {Kumar}(2020)}]{Beniamini&Kumar2020}
{Beniamini}, P., \& {Kumar}, P. 2020, \mnras, 498, 651, \dodoi{10.1093/mnras/staa2489}

\bibitem[{{Blaes} {et~al.}(1989){Blaes}, {Blandford}, {Goldreich}, \& {Madau}}]{Blaes1989}
{Blaes}, O., {Blandford}, R., {Goldreich}, P., \& {Madau}, P. 1989, \apj, 343, 839, \dodoi{10.1086/167754}

\bibitem[{{Bochenek} {et~al.}(2020){Bochenek}, {Ravi}, {Belov}, {Hallinan}, {Kocz}, {Kulkarni}, \& {McKenna}}]{Bochenek2020}
{Bochenek}, C.~D., {Ravi}, V., {Belov}, K.~V., {et~al.} 2020, \nat, 587, 59, \dodoi{10.1038/s41586-020-2872-x}

\bibitem[{{Bransgrove} {et~al.}(2020){Bransgrove}, {Beloborodov}, \& {Levin}}]{Bransgrove2020}
{Bransgrove}, A., {Beloborodov}, A.~M., \& {Levin}, Y. 2020, \apj, 897, 173, \dodoi{10.3847/1538-4357/ab93b7}

\bibitem[{{Chatterjee} {et~al.}(2017){Chatterjee}, {Law}, {Wharton}, {Burke-Spolaor}, {Hessels}, {Bower}, {Cordes}, {Tendulkar}, {Bassa}, {Demorest}, {Butler}, {Seymour}, {Scholz}, {Abruzzo}, {Bogdanov}, {Kaspi}, {Keimpema}, {Lazio}, {Marcote}, {McLaughlin}, {Paragi}, {Ransom}, {Rupen}, {Spitler}, \& {van Langevelde}}]{Chatterjee2017}
{Chatterjee}, S., {Law}, C.~J., {Wharton}, R.~S., {et~al.} 2017, \nat, 541, 58, \dodoi{10.1038/nature20797}

\bibitem[{{Chen} {et~al.}(2024){Chen}, {Yuan}, \& {Bernardi}}]{Chen2024}
{Chen}, A.~Y., {Yuan}, Y., \& {Bernardi}, D. 2024, arXiv e-prints, arXiv:2404.06431, \dodoi{10.48550/arXiv.2404.06431}

\bibitem[{{Chen} {et~al.}(2022){Chen}, {Yuan}, {Li}, \& {Mahlmann}}]{ChenYR2022}
{Chen}, A.~Y., {Yuan}, Y., {Li}, X., \& {Mahlmann}, J.~F. 2022, arXiv e-prints, arXiv:2210.13506, \dodoi{10.48550/arXiv.2210.13506}

\bibitem[{{CHIME/FRB Collaboration} {et~al.}(2020){CHIME/FRB Collaboration}, {Andersen}, {Bandura}, {Bhardwaj}, {Bij}, {Boyce}, {Boyle}, {Brar}, {Cassanelli}, {Chawla}, {Chen}, {Cliche}, {Cook}, {Cubranic}, {Curtin}, {Denman}, {Dobbs}, {Dong}, {Fandino}, {Fonseca}, {Gaensler}, {Giri}, {Good}, {Halpern}, {Hill}, {Hinshaw}, {H{\"o}fer}, {Josephy}, {Kania}, {Kaspi}, {Landecker}, {Leung}, {Li}, {Lin}, {Masui}, {McKinven}, {Mena-Parra}, {Merryfield}, {Meyers}, {Michilli}, {Milutinovic}, {Mirhosseini}, {M{\"u}nchmeyer}, {Naidu}, {Newburgh}, {Ng}, {Patel}, {Pen}, {Pinsonneault-Marotte}, {Pleunis}, {Quine}, {Rafiei-Ravandi}, {Rahman}, {Ransom}, {Renard}, {Sanghavi}, {Scholz}, {Shaw}, {Shin}, {Siegel}, {Singh}, {Smegal}, {Smith}, {Stairs}, {Tan}, {Tendulkar}, {Tretyakov}, {Vanderlinde}, {Wang}, {Wulf}, \& {Zwaniga}}]{CHIME/FRB2020}
{CHIME/FRB Collaboration}, {Andersen}, B.~C., {Bandura}, K.~M., {et~al.} 2020, \nat, 587, 54, \dodoi{10.1038/s41586-020-2863-y}

\bibitem[{{Chime/Frb Collabortion}(2021)}]{CHIME2021}
{Chime/Frb Collabortion}. 2021, The Astronomer's Telegram, 14497, 1

\bibitem[{{Cho} {et~al.}(2020){Cho}, {Macquart}, {Shannon}, {Deller}, {Morrison}, {Ekers}, {Bannister}, {Farah}, {Qiu}, {Sammons}, {Bailes}, {Bhandari}, {Day}, {James}, {Phillips}, {Prochaska}, \& {Tuthill}}]{Cho2020}
{Cho}, H., {Macquart}, J.-P., {Shannon}, R.~M., {et~al.} 2020, \apjl, 891, L38, \dodoi{10.3847/2041-8213/ab7824}

\bibitem[{{Cooper} \& {Wijers}(2021)}]{Cooper2021}
{Cooper}, A.~J., \& {Wijers}, R.~A.~M.~J. 2021, \mnras, 508, L32, \dodoi{10.1093/mnrasl/slab099}

\bibitem[{{Day} {et~al.}(2020){Day}, {Deller}, {Shannon}, {Qiu(邱昊)}, {Bannister}, {Bhandari}, {Ekers}, {Flynn}, {James}, {Macquart}, {Mahony}, {Phillips}, \& {Xavier Prochaska}}]{Day2020}
{Day}, C.~K., {Deller}, A.~T., {Shannon}, R.~M., {et~al.} 2020, \mnras, 497, 3335, \dodoi{10.1093/mnras/staa2138}

\bibitem[{{Feng} {et~al.}(2022{\natexlab{a}}){Feng}, {Zhang}, {Li}, {Yang}, {Wang}, {Niu}, {Dai}, \& {Yao}}]{Feng2022b}
{Feng}, Y., {Zhang}, Y.-K., {Li}, D., {et~al.} 2022{\natexlab{a}}, Science Bulletin, 67, 2398, \dodoi{10.1016/j.scib.2022.11.014}

\bibitem[{{Feng} {et~al.}(2022{\natexlab{b}}){Feng}, {Li}, {Yang}, {Zhang}, {Zhu}, {Zhang}, {Lu}, {Wang}, {Dai}, {Lynch}, {Yao}, {Jiang}, {Niu}, {Zhou}, {Xu}, {Miao}, {Niu}, {Meng}, {Qian}, {Tsai}, {Wang}, {Xue}, {Yue}, {Yuan}, {Zhang}, \& {Zhang}}]{Feng2022}
{Feng}, Y., {Li}, D., {Yang}, Y.-P., {et~al.} 2022{\natexlab{b}}, Science, 375, 1266, \dodoi{10.1126/science.abl7759}

\bibitem[{{Golbraikh} \& {Lyubarsky}(2023)}]{Golbraikh&Lyubarsky2023}
{Golbraikh}, E., \& {Lyubarsky}, Y. 2023, \apj, 957, 102, \dodoi{10.3847/1538-4357/acfa78}

\bibitem[{{Goldreich} \& {Julian}(1969)}]{Goldreich&Julian1969}
{Goldreich}, P., \& {Julian}, W.~H. 1969, \apj, 157, 869, \dodoi{10.1086/150119}

\bibitem[{{Herold}(1979)}]{Herold1979}
{Herold}, H. 1979, \prd, 19, 2868, \dodoi{10.1103/PhysRevD.19.2868}

\bibitem[{{Iwamoto} {et~al.}(2023){Iwamoto}, {Matsumoto}, {Amano}, {Matsukiyo}, \& {Hoshino}}]{Iwamoto2023}
{Iwamoto}, M., {Matsumoto}, Y., {Amano}, T., {Matsukiyo}, S., \& {Hoshino}, M. 2023, arXiv e-prints, arXiv:2311.18487, \dodoi{10.48550/arXiv.2311.18487}

\bibitem[{{Jackson}(1998)}]{Jackson1998}
{Jackson}, J.~D. 1998, {Classical Electrodynamics, 3rd Edition}

\bibitem[{{Jiang} {et~al.}(2022){Jiang}, {Wang}, {Xu}, {Xu}, {Zhang}, {Wang}, {Zhou}, {Zhang}, {Niu}, {Lee}, {Zhang}, {Han}, {Li}, {Zhu}, {Dai}, {Feng}, {Jing}, {Li}, {Luo}, {Miao}, {Niu}, {Tsai}, {Wang}, {Wang}, {Xu}, {Yang}, {Yang}, {Yao}, \& {Yuan}}]{Jiang22}
{Jiang}, J.-C., {Wang}, W.-Y., {Xu}, H., {et~al.} 2022, Research in Astronomy and Astrophysics, 22, 124003, \dodoi{10.1088/1674-4527/ac98f6}

\bibitem[{Kasahara(1981)}]{Kasahara1981}
Kasahara, K. 1981, Earthquake mechanics, Vol. 248 (Cambridge university press Cambridge)

\bibitem[{{Katz}(2018)}]{Katz2018}
{Katz}, J.~I. 2018, \mnras, 481, 2946, \dodoi{10.1093/mnras/sty2459}

\bibitem[{{Kumar} \& {Bo{\v{s}}njak}(2020)}]{kumar&Bosnjak2020}
{Kumar}, P., \& {Bo{\v{s}}njak}, {\v{Z}}. 2020, \mnras, 494, 2385, \dodoi{10.1093/mnras/staa774}

\bibitem[{{Kumar} {et~al.}(2017){Kumar}, {Lu}, \& {Bhattacharya}}]{Kumar2017}
{Kumar}, P., {Lu}, W., \& {Bhattacharya}, M. 2017, \mnras, 468, 2726, \dodoi{10.1093/mnras/stx665}

\bibitem[{{Kumar} {et~al.}(2024){Kumar}, {Qu}, \& {Zhang}}]{KQZ2024}
{Kumar}, P., {Qu}, Y., \& {Zhang}, B. 2024, arXiv e-prints, arXiv:2406.01266, \dodoi{10.48550/arXiv.2406.01266}

\bibitem[{{Li} {et~al.}(2021){Li}, {Lin}, {Xiong}, {Ge}, {Li}, {Li}, {Lu}, {Zhang}, {Tuo}, {Nang}, {Zhang}, {Xiao}, {Chen}, {Song}, {Xu}, {Liu}, {Jia}, {Cao}, {Qu}, {Zhang}, {Gu}, {Liao}, {Zhao}, {Tan}, {Nie}, {Zhao}, {Zheng}, {Zheng}, {Luo}, {Cai}, {Li}, {Xue}, {Bu}, {Chang}, {Chen}, {Chen}, {Chen}, {Chen}, {Chen}, {Cui}, {Cui}, {Deng}, {Dong}, {Du}, {Fu}, {Gao}, {Gao}, {Gao}, {Gu}, {Guan}, {Guo}, {Han}, {Huang}, {Huo}, {Jiang}, {Jiang}, {Jin}, {Jin}, {Kong}, {Li}, {Li}, {Li}, {Li}, {Li}, {Li}, {Li}, {Liang}, {Liu}, {Liu}, {Liu}, {Liu}, {Liu}, {Lu}, {Lu}, {Luo}, {Ma}, {Meng}, {Ou}, {Sai}, {Shang}, {Song}, {Sun}, {Tao}, {Wang}, {Wang}, {Wang}, {Wang}, {Wang}, {Wen}, {Wu}, {Wu}, {Wu}, {Xiao}, {Xu}, {Yang}, {Yang}, {Yang}, {Yang}, {Yi}, {Yin}, {You}, {Zhang}, {Zhang}, {Zhang}, {Zhang}, {Zhang}, {Zhang}, {Zhang}, {Zhang}, {Zhang}, {Zhou}, {Zhou}, {Zhu}, {Zhu}, \& {Zhuang}}]{CKLi21}
{Li}, C.~K., {Lin}, L., {Xiong}, S.~L., {et~al.} 2021, Nature Astronomy, 5, 378.
\newblock \doarXiv{2005.11071}

\bibitem[{{Liu} {et~al.}(2023){Liu}, {Wang}, {Yang}, \& {Dai}}]{Liu22}
{Liu}, Z.-N., {Wang}, W.-Y., {Yang}, Y.-P., \& {Dai}, Z.-G. 2023, \apj, 943, 47, \dodoi{10.3847/1538-4357/acac23}

\bibitem[{{Lorimer} {et~al.}(2007){Lorimer}, {Bailes}, {McLaughlin}, {Narkevic}, \& {Crawford}}]{Lorimer2007}
{Lorimer}, D.~R., {Bailes}, M., {McLaughlin}, M.~A., {Narkevic}, D.~J., \& {Crawford}, F. 2007, Science, 318, 777, \dodoi{10.1126/science.1147532}

\bibitem[{{Lu} {et~al.}(2020{\natexlab{a}}){Lu}, {Kumar}, \& {Zhang}}]{Lu20}
{Lu}, W., {Kumar}, P., \& {Zhang}, B. 2020{\natexlab{a}}, \mnras, 498, 1397, \dodoi{10.1093/mnras/staa2450}

\bibitem[{{Lu} {et~al.}(2020{\natexlab{b}}){Lu}, {Kumar}, \& {Zhang}}]{Lu2020}
---. 2020{\natexlab{b}}, \mnras, 498, 1397, \dodoi{10.1093/mnras/staa2450}

\bibitem[{{Luo} {et~al.}(2020){Luo}, {Wang}, {Men}, {Zhang}, {Jiang}, {Xu}, {Wang}, {Lee}, {Han}, {Zhang}, {Caballero}, {Chen}, {Chen}, {Gan}, {Guo}, {Hao}, {Huang}, {Jiang}, {Li}, {Li}, {Li}, {Luo}, {Pan}, {Pei}, {Qian}, {Sun}, {Wang}, {Wang}, {Wen}, {Xu}, {Xu}, {Yan}, {Yan}, {Yu}, {Yuan}, {Zhang}, \& {Zhu}}]{Luo2020nature}
{Luo}, R., {Wang}, B.~J., {Men}, Y.~P., {et~al.} 2020, \nat, 586, 693, \dodoi{10.1038/s41586-020-2827-2}

\bibitem[{{Lyubarsky}(2014)}]{Lyubarsky2014}
{Lyubarsky}, Y. 2014, \mnras, 442, L9, \dodoi{10.1093/mnrasl/slu046}

\bibitem[{{Lyubarsky}(2021)}]{Lyubarsky2021}
---. 2021, Universe, 7, 56, \dodoi{10.3390/universe7030056}

\bibitem[{{Lyutikov}(2021)}]{Lyutikov2021}
{Lyutikov}, M. 2021, \apj, 922, 166, \dodoi{10.3847/1538-4357/ac1b32}

\bibitem[{{Lyutikov}(2024)}]{Lyutikov2024}
---. 2024, \mnras, 529, 2180, \dodoi{10.1093/mnras/stae591}

\bibitem[{{Macquart} {et~al.}(2020){Macquart}, {Prochaska}, {McQuinn}, {Bannister}, {Bhandari}, {Day}, {Deller}, {Ekers}, {James}, {Marnoch}, {Os{\l}owski}, {Phillips}, {Ryder}, {Scott}, {Shannon}, \& {Tejos}}]{Macquart2020}
{Macquart}, J.~P., {Prochaska}, J.~X., {McQuinn}, M., {et~al.} 2020, \nat, 581, 391, \dodoi{10.1038/s41586-020-2300-2}

\bibitem[{{Mahlmann} {et~al.}(2024){Mahlmann}, {Aloy}, \& {Li}}]{Mahlmann2024}
{Mahlmann}, J.~F., {Aloy}, M.~{\'A}., \& {Li}, X. 2024, arXiv e-prints, arXiv:2405.12272, \dodoi{10.48550/arXiv.2405.12272}

\bibitem[{{Marcote} {et~al.}(2020){Marcote}, {Nimmo}, {Hessels}, {Tendulkar}, {Bassa}, {Paragi}, {Keimpema}, {Bhardwaj}, {Karuppusamy}, {Kaspi}, {Law}, {Michilli}, {Aggarwal}, {Andersen}, {Archibald}, {Bandura}, {Bower}, {Boyle}, {Brar}, {Burke-Spolaor}, {Butler}, {Cassanelli}, {Chawla}, {Demorest}, {Dobbs}, {Fonseca}, {Giri}, {Good}, {Gourdji}, {Josephy}, {Kirichenko}, {Kirsten}, {Landecker}, {Lang}, {Lazio}, {Li}, {Lin}, {Linford}, {Masui}, {Mena-Parra}, {Naidu}, {Ng}, {Patel}, {Pen}, {Pleunis}, {Rafiei-Ravandi}, {Rahman}, {Renard}, {Scholz}, {Siegel}, {Smith}, {Stairs}, {Vanderlinde}, \& {Zwaniga}}]{Marcote2020}
{Marcote}, B., {Nimmo}, K., {Hessels}, J.~W.~T., {et~al.} 2020, \nat, 577, 190, \dodoi{10.1038/s41586-019-1866-z}

\bibitem[{{Margalit} {et~al.}(2020){Margalit}, {Metzger}, \& {Sironi}}]{Margalit2020}
{Margalit}, B., {Metzger}, B.~D., \& {Sironi}, L. 2020, \mnras, 494, 4627, \dodoi{10.1093/mnras/staa1036}

\bibitem[{{Masui} {et~al.}(2015){Masui}, {Lin}, {Sievers}, {Anderson}, {Chang}, {Chen}, {Ganguly}, {Jarvis}, {Kuo}, {Li}, {Liao}, {McLaughlin}, {Pen}, {Peterson}, {Roman}, {Timbie}, {Voytek}, \& {Yadav}}]{masui2015}
{Masui}, K., {Lin}, H.-H., {Sievers}, J., {et~al.} 2015, \nat, 528, 523, \dodoi{10.1038/nature15769}

\bibitem[{{Mckinven} {et~al.}(2024){Mckinven}, {Bhardwaj}, {Eftekhari}, {Kilpatrick}, {Kirichenko}, {Pal}, {Cook}, {Gaensler}, {Giri}, {Kaspi}, {Michilli}, {Nimmo}, {Pearlman}, {Pleunis}, {Sand}, {Stairs}, {Andersen}, {Andrew}, {Bandura}, {Brar}, {Cassanelli}, {Chatterjee}, {Curtin}, {Dong}, {Eadie}, {Fonseca}, {Ibik}, {Kaczmarek}, {Kharel}, {Lazda}, {Leung}, {Li}, {Main}, {Masui}, {Mena-Parra}, {Ng}, {Pandhi}, {Shivraj Patil}, {Prochaska}, {Rafiei-Ravandi}, {Scholz}, {Shah}, {Shin}, \& {Smith}}]{Mckinven2024}
{Mckinven}, R., {Bhardwaj}, M., {Eftekhari}, T., {et~al.} 2024, arXiv e-prints, arXiv:2402.09304, \dodoi{10.48550/arXiv.2402.09304}

\bibitem[{{Melrose}(1978)}]{Melrose1978}
{Melrose}, D.~B. 1978, \apj, 225, 557, \dodoi{10.1086/156516}

\bibitem[{{Melrose} \& {McPhedran}(1991)}]{Melrose&McPhedran1991}
{Melrose}, D.~B., \& {McPhedran}, R.~C. 1991, {Electromagnetic Processes in Dispersive Media}

\bibitem[{{Mereghetti} {et~al.}(2020){Mereghetti}, {Savchenko}, {Ferrigno}, {G{\"o}tz}, {Rigoselli}, {Tiengo}, {Bazzano}, {Bozzo}, {Coleiro}, {Courvoisier}, {Doyle}, {Goldwurm}, {Hanlon}, {Jourdain}, {von Kienlin}, {Lutovinov}, {Martin-Carrillo}, {Molkov}, {Natalucci}, {Onori}, {Panessa}, {Rodi}, {Rodriguez}, {S{\'a}nchez-Fern{\'a}ndez}, {Sunyaev}, \& {Ubertini}}]{Mereghetti20}
{Mereghetti}, S., {Savchenko}, V., {Ferrigno}, C., {et~al.} 2020, \apjl, 898, L29, \dodoi{10.3847/2041-8213/aba2cf}

\bibitem[{{Metzger} {et~al.}(2019){Metzger}, {Margalit}, \& {Sironi}}]{Metzger2019}
{Metzger}, B.~D., {Margalit}, B., \& {Sironi}, L. 2019, \mnras, 485, 4091, \dodoi{10.1093/mnras/stz700}

\bibitem[{{Nimmo} {et~al.}(2022){Nimmo}, {Hessels}, {Kirsten}, {Keimpema}, {Cordes}, {Snelders}, {Hewitt}, {Karuppusamy}, {Archibald}, {Bezrukovs}, {Bhardwaj}, {Blaauw}, {Buttaccio}, {Cassanelli}, {Conway}, {Corongiu}, {Feiler}, {Fonseca}, {Forss{\'e}n}, {Gawro{\'n}ski}, {Giroletti}, {Kharinov}, {Leung}, {Lindqvist}, {Maccaferri}, {Marcote}, {Masui}, {Mckinven}, {Melnikov}, {Michilli}, {Mikhailov}, {Ng}, {Orbidans}, {Ould-Boukattine}, {Paragi}, {Pearlman}, {Petroff}, {Rahman}, {Scholz}, {Shin}, {Smith}, {Stairs}, {Surcis}, {Tendulkar}, {Vlemmings}, {Wang}, {Yang}, \& {Yuan}}]{Nimmo2022}
{Nimmo}, K., {Hessels}, J.~W.~T., {Kirsten}, F., {et~al.} 2022, Nature Astronomy, 6, 393, \dodoi{10.1038/s41550-021-01569-9}

\bibitem[{{Nishiura} \& {Ioka}(2024)}]{Nishiura&Ioka2024}
{Nishiura}, R., \& {Ioka}, K. 2024, \prd, 109, 043048, \dodoi{10.1103/PhysRevD.109.043048}

\bibitem[{{Niu} {et~al.}(2022{\natexlab{a}}){Niu}, {Aggarwal}, {Li}, {Zhang}, {Chatterjee}, {Tsai}, {Yu}, {Law}, {Burke-Spolaor}, {Cordes}, {Zhang}, {Ocker}, {Yao}, {Wang}, {Feng}, {Niino}, {Bochenek}, {Cruces}, {Connor}, {Jiang}, {Dai}, {Luo}, {Li}, {Miao}, {Niu}, {Anna-Thomas}, {Sydnor}, {Stern}, {Wang}, {Yuan}, {Yue}, {Zhou}, {Yan}, {Zhu}, \& {Zhang}}]{Niu2022}
{Niu}, C.~H., {Aggarwal}, K., {Li}, D., {et~al.} 2022{\natexlab{a}}, \nat, 606, 873, \dodoi{10.1038/s41586-022-04755-5}

\bibitem[{{Niu} {et~al.}(2022{\natexlab{b}}){Niu}, {Zhu}, {Zhang}, {Yuan}, {Zhou}, {Zhang}, {Jiang}, {Han}, {Li}, {Lee}, {Wang}, {Feng}, {Li}, {Luo}, {Wang}, {Dai}, {Miao}, {Niu}, {Xu}, {Zhang}, {Wang}, {Wang}, \& {Xu}}]{NiuJR2022}
{Niu}, J.-R., {Zhu}, W.-W., {Zhang}, B., {et~al.} 2022{\natexlab{b}}, Research in Astronomy and Astrophysics, 22, 124004, \dodoi{10.1088/1674-4527/ac995d}

\bibitem[{{Petroff} {et~al.}(2019){Petroff}, {Hessels}, \& {Lorimer}}]{petroff2019}
{Petroff}, E., {Hessels}, J.~W.~T., \& {Lorimer}, D.~R. 2019, \aapr, 27, 4, \dodoi{10.1007/s00159-019-0116-6}

\bibitem[{{Petroff} {et~al.}(2015){Petroff}, {Bailes}, {Barr}, {Barsdell}, {Bhat}, {Bian}, {Burke-Spolaor}, {Caleb}, {Champion}, {Chandra}, {Da Costa}, {Delvaux}, {Flynn}, {Gehrels}, {Greiner}, {Jameson}, {Johnston}, {Kasliwal}, {Keane}, {Keller}, {Kocz}, {Kramer}, {Leloudas}, {Malesani}, {Mulchaey}, {Ng}, {Ofek}, {Perley}, {Possenti}, {Schmidt}, {Shen}, {Stappers}, {Tisserand}, {van Straten}, \& {Wolf}}]{petroff2015}
{Petroff}, E., {Bailes}, M., {Barr}, E.~D., {et~al.} 2015, \mnras, 447, 246, \dodoi{10.1093/mnras/stu2419}

\bibitem[{{Pleunis} {et~al.}(2021){Pleunis}, {Good}, {Kaspi}, {Mckinven}, {Ransom}, {Scholz}, {Bandura}, {Bhardwaj}, {Boyle}, {Brar}, {Cassanelli}, {Chawla}, {(Adam) Dong}, {Fonseca}, {Gaensler}, {Josephy}, {Kaczmarek}, {Leung}, {Lin}, {Masui}, {Mena-Parra}, {Michilli}, {Ng}, {Patel}, {Rafiei-Ravandi}, {Rahman}, {Sanghavi}, {Shin}, {Smith}, {Stairs}, \& {Tendulkar}}]{Pleunis2021}
{Pleunis}, Z., {Good}, D.~C., {Kaspi}, V.~M., {et~al.} 2021, \apj, 923, 1, \dodoi{10.3847/1538-4357/ac33ac}

\bibitem[{{Plotnikov} \& {Sironi}(2019)}]{Plotnikov&Sironi2019}
{Plotnikov}, I., \& {Sironi}, L. 2019, \mnras, 485, 3816, \dodoi{10.1093/mnras/stz640}

\bibitem[{{Prochaska} {et~al.}(2019){Prochaska}, {Macquart}, {McQuinn}, {Simha}, {Shannon}, {Day}, {Marnoch}, {Ryder}, {Deller}, {Bannister}, {Bhandari}, {Bordoloi}, {Bunton}, {Cho}, {Flynn}, {Mahony}, {Phillips}, {Qiu}, \& {Tejos}}]{Prochaska2019}
{Prochaska}, J.~X., {Macquart}, J.-P., {McQuinn}, M., {et~al.} 2019, Science, 366, 231, \dodoi{10.1126/science.aay0073}

\bibitem[{{Qu} {et~al.}(2022){Qu}, {Kumar}, \& {Zhang}}]{QKZ}
{Qu}, Y., {Kumar}, P., \& {Zhang}, B. 2022, \mnras, 515, 2020, \dodoi{10.1093/mnras/stac1910}

\bibitem[{{Qu} \& {Zhang}(2023)}]{Qu&Zhang2023}
{Qu}, Y., \& {Zhang}, B. 2023, \mnras, 522, 2448, \dodoi{10.1093/mnras/stad1072}

\bibitem[{{Qu} {et~al.}(2023){Qu}, {Zhang}, \& {Kumar}}]{QZK}
{Qu}, Y., {Zhang}, B., \& {Kumar}, P. 2023, \mnras, 518, 66, \dodoi{10.1093/mnras/stac3111}

\bibitem[{{Ravi} {et~al.}(2019){Ravi}, {Catha}, {D'Addario}, {Djorgovski}, {Hallinan}, {Hobbs}, {Kocz}, {Kulkarni}, {Shi}, {Vedantham}, {Weinreb}, \& {Woody}}]{Ravi2019}
{Ravi}, V., {Catha}, M., {D'Addario}, L., {et~al.} 2019, \nat, 572, 352, \dodoi{10.1038/s41586-019-1389-7}

\bibitem[{{Ruderman} \& {Sutherland}(1975)}]{Ruderman1975}
{Ruderman}, M.~A., \& {Sutherland}, P.~G. 1975, \apj, 196, 51, \dodoi{10.1086/153393}

\bibitem[{{Rybicki} \& {Lightman}(1979)}]{Rybicki&Lightman1979}
{Rybicki}, G.~B., \& {Lightman}, A.~P. 1979, {Radiative processes in astrophysics}

\bibitem[{{Sheikh} {et~al.}(2024){Sheikh}, {Farah}, {Pollak}, {Siemion}, {Chamma}, {Cruz}, {Davis}, {DeBoer}, {Gajjar}, {Karn}, {Kittling}, {Lu}, {Masters}, {Premnath}, {Schoultz}, {Shumaker}, {Singh}, \& {Snodgrass}}]{Sheikh2024}
{Sheikh}, S.~Z., {Farah}, W., {Pollak}, A.~W., {et~al.} 2024, \mnras, 527, 10425, \dodoi{10.1093/mnras/stad3630}

\bibitem[{{Sherman} {et~al.}(2023){Sherman}, {Connor}, {Ravi}, {Law}, {Chen}, {Catha}, {Faber}, {Hallinan}, {Harnach}, {Hellbourg}, {Hobbs}, {Hodge}, {Hodges}, {Lamb}, {Rasmussen}, {Sharma}, {Shi}, {Simard}, {Somalwar}, {Squillace}, {Weinreb}, {Woody}, \& {Yadlapalli}}]{Sherman2023}
{Sherman}, M.~B., {Connor}, L., {Ravi}, V., {et~al.} 2023, arXiv e-prints, arXiv:2308.06813, \dodoi{10.48550/arXiv.2308.06813}

\bibitem[{{Snelders} {et~al.}(2023){Snelders}, {Nimmo}, {Hessels}, {Bensellam}, {Zwaan}, {Chawla}, {Ould-Boukattine}, {Kirsten}, {Faber}, \& {Gajjar}}]{Snelders2023}
{Snelders}, M.~P., {Nimmo}, K., {Hessels}, J.~W.~T., {et~al.} 2023, Nature Astronomy, \dodoi{10.1038/s41550-023-02101-x}

\bibitem[{{Spitler} {et~al.}(2014){Spitler}, {Cordes}, {Hessels}, {Lorimer}, {McLaughlin}, {Chatterjee}, {Crawford}, {Deneva}, {Kaspi}, {Wharton}, {Allen}, {Bogdanov}, {Brazier}, {Camilo}, {Freire}, {Jenet}, {Karako-Argaman}, {Knispel}, {Lazarus}, {Lee}, {van Leeuwen}, {Lynch}, {Ransom}, {Scholz}, {Siemens}, {Stairs}, {Stovall}, {Swiggum}, {Venkataraman}, {Zhu}, {Aulbert}, \& {Fehrmann}}]{Spitler2014}
{Spitler}, L.~G., {Cordes}, J.~M., {Hessels}, J.~W.~T., {et~al.} 2014, \apj, 790, 101, \dodoi{10.1088/0004-637X/790/2/101}

\bibitem[{{Tendulkar} {et~al.}(2017){Tendulkar}, {Bassa}, {Cordes}, {Bower}, {Law}, {Chatterjee}, {Adams}, {Bogdanov}, {Burke-Spolaor}, {Butler}, {Demorest}, {Hessels}, {Kaspi}, {Lazio}, {Maddox}, {Marcote}, {McLaughlin}, {Paragi}, {Ransom}, {Scholz}, {Seymour}, {Spitler}, {van Langevelde}, \& {Wharton}}]{Tendulkar2017}
{Tendulkar}, S.~P., {Bassa}, C.~G., {Cordes}, J.~M., {et~al.} 2017, \apjl, 834, L7, \dodoi{10.3847/2041-8213/834/2/L7}

\bibitem[{{Thornton} {et~al.}(2013){Thornton}, {Stappers}, {Bailes}, {Barsdell}, {Bates}, {Bhat}, {Burgay}, {Burke-Spolaor}, {Champion}, {Coster}, {D'Amico}, {Jameson}, {Johnston}, {Keith}, {Kramer}, {Levin}, {Milia}, {Ng}, {Possenti}, \& {van Straten}}]{Thornton2013}
{Thornton}, D., {Stappers}, B., {Bailes}, M., {et~al.} 2013, Science, 341, 53, \dodoi{10.1126/science.1236789}

\bibitem[{{Wadiasingh} \& {Timokhin}(2019)}]{Wadiasingh19}
{Wadiasingh}, Z., \& {Timokhin}, A. 2019, \apj, 879, 4, \dodoi{10.3847/1538-4357/ab2240}

\bibitem[{{Wang} {et~al.}(2019){Wang}, {Zhang}, {Chen}, \& {Xu}}]{Wang2019}
{Wang}, W., {Zhang}, B., {Chen}, X., \& {Xu}, R. 2019, \apjl, 876, L15, \dodoi{10.3847/2041-8213/ab1aab}

\bibitem[{{Wang} {et~al.}(2022{\natexlab{a}}){Wang}, {Jiang}, {Lee}, {Xu}, \& {Zhang}}]{Wang2022b}
{Wang}, W.-Y., {Jiang}, J.-C., {Lee}, K., {Xu}, R., \& {Zhang}, B. 2022{\natexlab{a}}, \mnras, 517, 5080, \dodoi{10.1093/mnras/stac3070}

\bibitem[{{Wang} {et~al.}(2023){Wang}, {Yang}, {Li}, {Liu}, \& {Xu}}]{Wang2023}
{Wang}, W.-Y., {Yang}, Y.-P., {Li}, H.-B., {Liu}, J., \& {Xu}, R. 2023, arXiv e-prints, arXiv:2311.13114, \dodoi{10.48550/arXiv.2311.13114}

\bibitem[{{Wang} {et~al.}(2022{\natexlab{b}}){Wang}, {Yang}, {Niu}, {Xu}, \& {Zhang}}]{Wang2022}
{Wang}, W.-Y., {Yang}, Y.-P., {Niu}, C.-H., {Xu}, R., \& {Zhang}, B. 2022{\natexlab{b}}, \apj, 927, 105, \dodoi{10.3847/1538-4357/ac4097}

\bibitem[{{Wang} {et~al.}(2024){Wang}, {Zhang}, {Zhou}, {Liu}, {Niu}, {Zhou}, {Gao}, {Liu}, {Xu}, \& {Zhang}}]{Wang2024}
{Wang}, W.-Y., {Zhang}, C., {Zhou}, E., {et~al.} 2024, arXiv e-prints, arXiv:2405.07152, \dodoi{10.48550/arXiv.2405.07152}

\bibitem[{{Xu} {et~al.}(2022){Xu}, {Niu}, {Chen}, {Lee}, {Zhu}, {Dong}, {Zhang}, {Jiang}, {Wang}, {Xu}, {Zhang}, {Fu}, {Filippenko}, {Peng}, {Zhou}, {Zhang}, {Wang}, {Feng}, {Li}, {Brink}, {Li}, {Lu}, {Yang}, {Caballero}, {Cai}, {Chen}, {Dai}, {Djorgovski}, {Esamdin}, {Gan}, {Guhathakurta}, {Han}, {Hao}, {Huang}, {Jiang}, {Li}, {Li}, {Li}, {Li}, {Li}, {Liu}, {Luo}, {Men}, {Niu}, {Peng}, {Qian}, {Song}, {Stern}, {Stockton}, {Sun}, {Wang}, {Wang}, {Wang}, {Wang}, {Wu}, {Xiao}, {Xiong}, {Xu}, {Xu}, {Yang}, {Yang}, {Yao}, {Yi}, {Yue}, {Yu}, {Yu}, {Yuan}, {Zhang}, {Zhang}, {Zhang}, {Zhao}, {Zheng}, {Zhu}, \& {Zou}}]{Xu2021}
{Xu}, H., {Niu}, J.~R., {Chen}, P., {et~al.} 2022, \nat, 609, 685, \dodoi{10.1038/s41586-022-05071-8}

\bibitem[{{Yang}(2023)}]{Yang2023}
{Yang}, Y.-P. 2023, \apj, 956, 67, \dodoi{10.3847/1538-4357/acebc6}

\bibitem[{{Yang} \& {Zhang}(2018)}]{Yang&zhang2018}
{Yang}, Y.-P., \& {Zhang}, B. 2018, \apj, 868, 31, \dodoi{10.3847/1538-4357/aae685}

\bibitem[{{Yang} \& {Zhang}(2021)}]{Yang&zhang2021}
---. 2021, \apj, 919, 89, \dodoi{10.3847/1538-4357/ac14b5}

\bibitem[{{Yang} \& {Zhang}(2023)}]{Yang&Zhang2023}
---. 2023, \mnras, 522, 4907, \dodoi{10.1093/mnras/stad1311}

\bibitem[{{Yang} {et~al.}(2020){Yang}, {Zhu}, {Zhang}, \& {Wu}}]{Yang2020}
{Yang}, Y.-P., {Zhu}, J.-P., {Zhang}, B., \& {Wu}, X.-F. 2020, \apjl, 901, L13, \dodoi{10.3847/2041-8213/abb535}

\bibitem[{{Yuan} {et~al.}(2022){Yuan}, {Beloborodov}, {Chen}, {Levin}, {Most}, \& {Philippov}}]{Yuan2022}
{Yuan}, Y., {Beloborodov}, A.~M., {Chen}, A.~Y., {et~al.} 2022, \apj, 933, 174, \dodoi{10.3847/1538-4357/ac7529}

\bibitem[{{Yuan} {et~al.}(2021){Yuan}, {Levin}, {Bransgrove}, \& {Philippov}}]{Yuan2021}
{Yuan}, Y., {Levin}, Y., {Bransgrove}, A., \& {Philippov}, A. 2021, \apj, 908, 176, \dodoi{10.3847/1538-4357/abd405}

\bibitem[{{Zhang}(2020)}]{Zhang2020}
{Zhang}, B. 2020, \nat, 587, 45, \dodoi{10.1038/s41586-020-2828-1}

\bibitem[{{Zhang}(2022)}]{Zhang22}
---. 2022, \apj, 925, 53, \dodoi{10.3847/1538-4357/ac3979}

\bibitem[{{Zhang}(2023{\natexlab{a}})}]{ZhangRMP}
---. 2023{\natexlab{a}}, Reviews of Modern Physics, 95, 035005, \dodoi{10.1103/RevModPhys.95.035005}

\bibitem[{{Zhang}(2023{\natexlab{b}})}]{Zhang2023Universe}
---. 2023{\natexlab{b}}, Universe, 9, 375, \dodoi{10.3390/universe9080375}

\bibitem[{{Zhang} {et~al.}(2023){Zhang}, {Li}, {Zhang}, {Cao}, {Feng}, {Wang}, {Qu}, {Niu}, {Zhu}, {Han}, {Jiang}, {Lee}, {Li}, {Luo}, {Niu}, {Tsai}, {Wang}, {Wang}, {Wu}, {Xu}, {Yang}, {Zhang}, {Zhou}, \& {Zhu}}]{ZhangYK2023}
{Zhang}, Y.-K., {Li}, D., {Zhang}, B., {et~al.} 2023, \apj, 955, 142, \dodoi{10.3847/1538-4357/aced0b}

\bibitem[{{Zhou} {et~al.}(2022){Zhou}, {Han}, {Zhang}, {Lee}, {Zhu}, {Li}, {Jing}, {Wang}, {Zhang}, {Jiang}, {Niu}, {Luo}, {Xu}, {Zhang}, {Wang}, {Xu}, {Wang}, {Yang}, \& {Feng}}]{ZhouDJ2022}
{Zhou}, D.~J., {Han}, J.~L., {Zhang}, B., {et~al.} 2022, Research in Astronomy and Astrophysics, 22, 124001, \dodoi{10.1088/1674-4527/ac98f8}

\end{thebibliography}

\end{document}